\documentclass[pra,preprintnumbers,twocolumn,amsmath,amssymb,superscriptaddress]{revtex4}
\usepackage{color}
\usepackage{graphicx}% Include figure files
\usepackage{dcolumn}% Align table columns on decimal point
\usepackage{bm}% bold math
\usepackage{hyperref}
\usepackage{mathrsfs}
\usepackage{soul}
\usepackage{ulem}
\usepackage{multirow}

\usepackage{braket}

\newcommand{\ie}{\textit{i.e.} }

\pdfoutput=1

\begin{document}

\title{Quantum teleportation of cat states with binary-outcome measurements}

\author{Jingyan Feng}
\thanks{These authors contributed equally}
\thanks{Corresponding author: {jf161@nyu.edu}}
\affiliation{State Key Laboratory of Precision Spectroscopy, School of Physical and Material Sciences, East China Normal University, Shanghai 200062, China}
\affiliation{New York University Shanghai; NYU-ECNU Institute of Physics at NYU Shanghai, 567 West Yangsi Road, Shanghai, 200124, China.}

\author{Mohan Zhang}
\thanks{These authors contributed equally}
\affiliation{State Key Laboratory of Precision Spectroscopy, School of Physical and Material Sciences, East China Normal University, Shanghai 200062, China}

\author{Matteo Fadel}
\thanks{Corresponding author: fadelm@phys.ethz.ch}
\affiliation{Department of Physics, ETH Zürich, 8093 Zürich, Switzerland}

\author{Tim Byrnes}
\thanks{Corresponding author: tim.byrnes@nyu.edu}
\affiliation{New York University Shanghai; NYU-ECNU Institute of Physics at NYU Shanghai, 567 West Yangsi Road, Shanghai, 200124, China.}
\affiliation{State Key Laboratory of Precision Spectroscopy, School of Physical and Material Sciences, East China Normal University, Shanghai 200062, China}
\affiliation{Center for Quantum and Topological Systems (CQTS), NYUAD Research Institute, New York University Abu Dhabi, UAE.}
\affiliation{Department of Physics, New York University, New York, NY 10003, USA}

\begin{abstract}
We propose a teleportation protocol involving beam splitting operations and binary-outcome measurements, such as parity measurements. These operations have a straightforward implementation using the dispersive regime of the Jaynes-Cummings Hamiltonian, making our protocol suitable for a broad class of platforms, including trapped ions, circuit quantum electrodynamics and acoustodynamics systems. In these platforms homodyne measurements of the bosonic modes are less natural than dispersive measurements, making standard continuous variable teleportation unsuitable. In our protocol, Alice is in possession of two bosonic modes and Bob a single mode. An entangled mode pair between Alice and Bob is created by performing a beam splitter operation on a cat state. An unknown qubit state encoded by cat states is then teleported from Alice to Bob after a beamsplitting operation, measurement sequence, and a conditional correction.  In the case of multiple measurements, near-perfect fidelity can be obtained.  We discuss the optimal parameters in order to maximize the fidelity under a variety of scenarios. 
\end{abstract}

\maketitle

\section{Introduction}

Quantum teleportation is one of the most striking applications of quantum entanglement, enabling the transfer of an unknown quantum state from one system to another using entanglement and feedforward operations \cite{Bennett93,Karlsson98}. It is a fundamental operation that serves as a cornerstone for numerous quantum protocols, underpinning advances in fields such as quantum communication and quantum computing. It plays a crucial role in enabling secure information transfer, distributed quantum networks, and the construction of scalable quantum architectures by facilitating state transfer without the need for direct physical transport of qubits. 

Since its initial experimental demonstration with optical photons in Ref. \cite{Bouwmeester97}, a wide range of quantum teleportation protocols have been proposed and implemented. The most basic version, involving qubit-based systems, has since been realized on numerous platforms \cite{Boschi98qubits,Barrett2004qubits,Xianmin10qubits,Nolleke13qubits}. The process relies on entangled pairs of qubits being utilized as a resource for transferring quantum states. In Ref. \cite{Ryan-Anderson2024}, the authors demonstrated fault-tolerant quantum state teleportation on a trapped-ion platform, marking significant progress toward universal quantum computing. In Ref. \cite{Valivarthi20prxq}, the authors demonstrated a teleportation fidelity of over 90\%, using time-bin encoded qubits. Recently, there has been increasing interest in extending teleportation to high-dimensional systems \cite{PhysRevLett.123.070505,PhysRevLett.125.230501,PhysRevA.100.032330,zhang2006experimental,Erhard20} and encoded systems \cite{huang2021emulating,luo2021quantum,chaudhary2024macroscopic,ryan2024high}.

In parallel to teleportation in discrete-variable systems, protocols for continuous-variable (CV) systems were also theoretically and experimentally developed \cite{Vaidman94CV,Braunstein98CV,Bowen03CV,Yukawa08CV,Shengshuai24CV}.  Here, the scheme typically involves sharing a CV-entangled state, such as a two-mode squeezed state, and quadrature measurements via homodyne detection are performed to transmit quantum information.  CV quantum teleportation was experimentally demonstrated for the first time by Furusawa and colleagues \cite{Furusawa98}, realizing unconditional teleportation. 
Subsequently, several variants of the CV teleportation were explored. In Ref.~\cite{Lee2011}, the authors proposed a hybrid approach combining discrete and continuous variable techniques. Moreover, CV quantum teleportation between atomic ensembles has been demonstrated \cite{Krauter2013}, and the teleportation of four degrees of freedom in a system of optical modes in Ref. \cite{Shengshuai24CV}.
These efforts motivate the exploration of quantum teleportation protocols in other systems.

In standard CV teleportation, quadrature measurements are a crucial step to transmit the unknown state from Alice to Bob. While this can be practically implemented experimentally in optical systems, for example, they are not the natural type of measurements performed in other systems such as  microwave cavity modes in circuit quantum electrodynamics (QED), vibration modes in circuit quantum acoustodynamics (QAD), or motional states of trapped ions. In fact, in the above systems the microwave modes or motional states are measured by coupling them to a two-level system through a Jaynes-Cummings (JC) interaction \cite{BlaisRMP21,von2022parity,LeibfriedRMP03}, resulting in the possibility of extracting only one bit of information per measurement repetition. For this reason, it is much easier and time-efficient to extract binary observables such as parity, rather than performing continuous variable measurements of phase-space quadratures. While alternative approaches to measure quadratures e.g. cQED exist, for example through parametrically-engineered longitudinal coupling or optimal control pulses \cite{Tanjung25}, parity measurements are still more robust and efficient due to their immediate implementation in the dispersive regime of the Jaynes-Cummings Hamiltonian. For this reason, having available a teleportation protocol which relies only on parity measurements is of central interest.

\begin{figure}[t]
    \centering
    \includegraphics[width=\columnwidth]{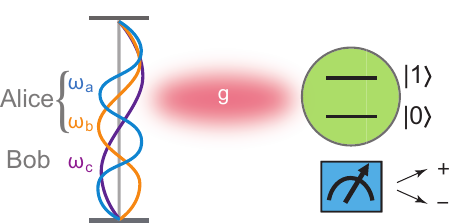}
    \caption{Schematic illustration of the physical system that is under consideration in this paper.  Three modes with frequencies $\omega_n$ couple to a qubit with strength $ g $. Modes $ a $ and $ b $ are in possession by Alice and mode $ c $ is in possession by Bob.  
 The qubit is read out in the $ \sigma^x $ eigenbasis, realizing the binary-outcome measurement.  This basic setup is used to realize the operations and measurements required for our teleportation protocol.}
    \label{fig:lilunhuitu}
\end{figure}

In this paper, we propose a protocol for quantum teleportation of cat states using beam splitters, displacement operations, and dispersive measurements such as parity measurements. The primary system that we have in mind is shown in Fig. \ref{fig:lilunhuitu}. A set of three individually addressable bosonic modes are coupled to a qubit which enables state preparation, multimode operations and measurement.  We consider two of the bosonic modes being in possession by Alice and the remaining mode by Bob. The aim is to teleport a quantum state that is on one of Alice's modes to Bob.  Using the scheme that is summarized in Fig. \ref{fig:circuit} (explained in more detail later), we demonstrate that high-fidelity teleportation can be achieved exclusively through operations permitted by the JC interaction.

This paper is organized as follows. In Sec. \ref{section2}, we introduce the experimental system and operations that we assume are available to perform the teleportation protocol. In Sec. \ref{section3}, we describe and analyze the entire quantum teleportation protocol analytically.  In Sec. \ref{singleparityresults}, we show numerical simulation results based on the single parity measurement version of the teleportation protocol.  In Sec. 
\ref{jointparityresults} we show the corresponding simulation results for joint parity measurements. Finally, in Sec. \ref{conclusion} we summarize and discuss our results.

\section{Experimental system and available operations}
\label{section2}

The experimental platform under consideration consists of a physical system with three bosonic modes and at least one qubit, interacting via a Jaynes-Cummings Hamiltonian  (Fig. \ref{fig:lilunhuitu}). Concrete examples of the physical system include circuit-QED \cite{BlaisRMP21} or QAD \cite{von2022parity,von2024engineering} where different microwave cavity modes or acoustic vibration modes are coupled to a superconducting qubit.  Alternatively, trapped ions \cite{LeibfriedRMP03} where motional degrees of freedom are coupled to internal (spin) states are another potential realization. In these systems, the oscillatory modes are treated as independent and are typically distinguished by their frequencies.

In the following, we discuss the operations and measurements that we assume are available on the bosonic modes.   In the next section, we will show how these are combined in order to realize our teleportation protocol.

\subsection{Preparation of states}

One of the operations that we assume is available is a phase-space displacement described by the operator
\begin{align}
    D(\alpha) = e^{\alpha a^\dagger - \alpha^* a } .  
\end{align}
Starting from the vacuum state  $ | 0 \rangle $, this generates a coherent state defined as 
\begin{align}	
|\alpha \rangle = D(\alpha) | 0 \rangle = e^{-|\alpha|^2 / 2} e^{\alpha a^\dagger} |0 \rangle .    
\end{align}
The amplitude and phase of the coherent state can be controlled by setting the appropriate amplitude and phase of the driving signal.

In our work, we will consider the teleportation in the basis of Schr\"odinger cat states, namely coherent superpositions of coherent states that take the form $\ket{\alpha}+e^{i\phi}\ket{-\alpha}$. These can be experimentally prepared using different approaches, such as through conditional-displacement gates \cite{Monroe1996, Vlastakis2013}, parity measurements \cite{Sun2014}, or nonlinear evolutions \cite{Kirchmair2013,Deléglise2008,bild2023schrodinger}.
As an example of the latter approach, it is possible to generate a cat state by interacting the mode with a qubit that follows the JC interaction 
\begin{align}
    H_{\text{JC}} = g(a^\dagger \sigma^- + a \sigma^+)
    \label{hamJC}
\end{align}
for a time $ t^\ast = \pi \alpha/g $, where the initial state is $|0 \rangle | \alpha \rangle $ and $g$ is the coupling strength.  Here $ \sigma^{\pm} = (\sigma^x \pm i \sigma^y )/2$ and $ \sigma^x,\sigma^y,\sigma^z $ are Pauli matrices. At time $t^\ast$ the qubit and mode disentangle, leaving the mode in (approximately) a Schr\"odinger cat state \cite{GeaCatPRA91}
\begin{align}
e^{-i H_{\text{JC}} \pi \alpha/g} |0 \rangle | \alpha \rangle \simeq |0 \rangle (| \alpha \rangle + | - \alpha \rangle )/\sqrt{2} .
\end{align}

\subsection{Mode transformations}

Teleportation protocols require operations between bosonic modes, which can also be implemented through a (multimode) JC Hamiltonian. In particular, the nonlinearity of an off-resonantly coupled qubit can be used to mediate interactions between different modes through parametric interactions. For example, effective beam-splitter (BS) transformations have been demonstrated between motional states of trapped ions \cite{Leibfried02}, microwave modes in cQED \cite{gao18} and acoustic vibration modes in cQAD \cite{von2024engineering}.
The resulting BS mode transformations are given by 
\begin{align}	
    a \rightarrow \frac{1}{\sqrt{2}} (a+b), \nonumber\\
    b\rightarrow \frac{1}{\sqrt{2}} (a-b).
    \label{beamsplitter}
\end{align}

\begin{figure}[t]
    \centering
    \includegraphics[width=\columnwidth]{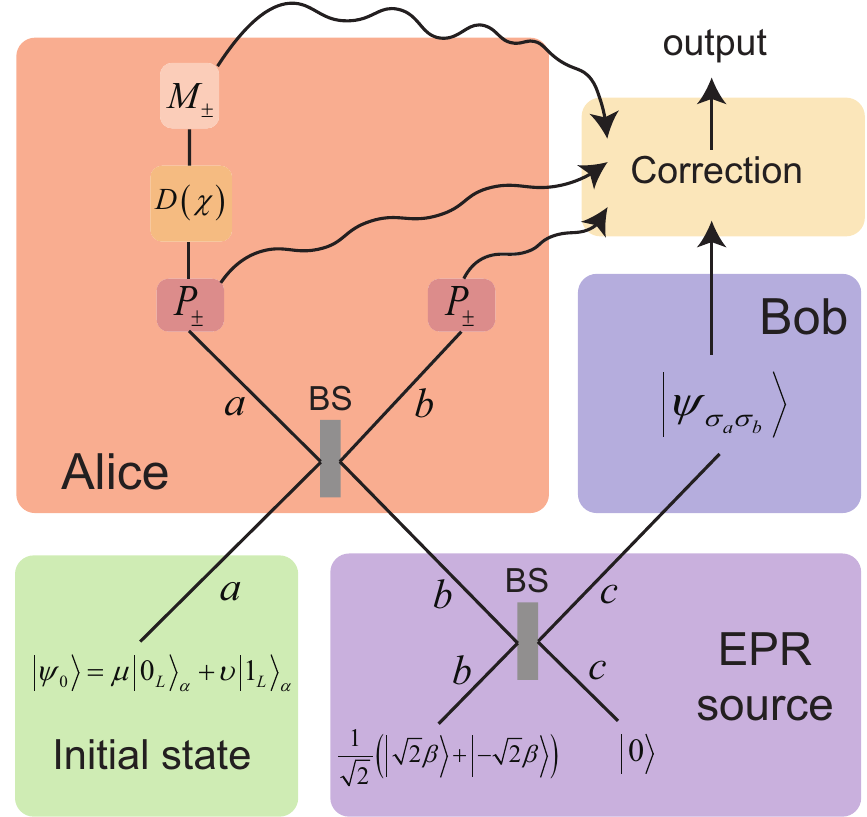}
    \caption{Our proposed quantum teleportation protocol. The sequence proceeds from bottom to top and is presented in a similar way to an optical circuit. We present our protocol as an optical circuit for the sake of visualization, and should not be viewed as a literal experimental configuration.   Straight lines represent the paths of the modes.   BS denotes a beam splitter, $ P_\pm $ denotes a parity measurement, $ D( \chi) $ is a displacement operation, and $ M_\pm $ is a dispersive measurement. The schematic shows the sequence using single-mode parity measurements (``Variant 1'').  For the protocol using joint parity measurements, these two measurements would be replaced by a common measurement between the two modes  (``Variant 2'').  Otherwise the two protocols are identical.   Straight lines denote state evolutions, while wiggly lines denote classical communication.   }
    \label{fig:circuit}
\end{figure}

\subsection{Measurement of modes}

By working in the strong dispersive regime of the JC Hamiltonian, it is possible to perform parity measurements of the mode \cite{LutterbachPRL97,Vlastakis2013,von2022parity}. The effective Hamiltonian for the single mode $f$ in the dispersive regime reads 
\begin{align}
H_{\text{disp}} \propto \omega f^\dagger f - G \sigma^z f^\dagger f ,  
\end{align}
where the phonon frequency $\omega $ is set to $ \frac{|g^2|}{\Delta}$, and $ G$ is the dispersive coupling parameter $G = \frac{|g|^2}{\Delta} $. 
Using the qubit as the measurement pointer, this Hamiltonian allows us to perform a measurement of the mode.  Initializing the qubit in the state $ | + \rangle = (|0 \rangle + |1 \rangle )/\sqrt{2} $, interacting the Hamiltonian for a time $ t $, and projecting the qubit in the $ \sigma^x $ eigenbasis leaves the modes in a state that is equivalent to applying the measurement operators \cite{mao2023measurement}
\begin{align}
M_+ (\tau) = \sum_{n} e^{-in\omega t} \cos (n \tau ) | n \rangle \langle n | \nonumber \\
M_- (\tau) = \sum_{n} e^{-in\omega t} \sin (n \tau) | n \rangle \langle n | ,
\label{dispmeas}
\end{align}
where $ | n \rangle = (a^\dagger)^n | 0 \rangle /\sqrt{n!} $ are the Fock states and the dimensionless interaction time is $ \tau = G t$.  The measurement operators satisfy the positive operator valued measure (POVM) relation $ M_+^\dagger M_+ + M_-^\dagger M_- = I $. 

Choosing the interaction time $ \tau = \pi/2 $ the two measurement outcomes reduce to a parity measurement
\begin{align}	
    P_{+} &  = \sum_{n \in \text{even}} |n\rangle \langle n|, \nonumber\\
    P_{-} &  = \sum_{n \in \text{odd}} |n\rangle \langle n|.
\end{align}

\section{Teleportation protocol}
\label{section3}

We now describe our quantum teleportation protocol. The overall scheme is summarized schematically in Fig. \ref{fig:circuit}. We draw the protocol in the style of a optical circuit, but it should be understood that this is a schematic representation, and does not constitute how the experiment physically appears. Alice is in possession of the two modes $ a,b$ and Bob has the mode $ c $, and the aim is to teleport the state initially on mode $ a $ to mode $ c $.  In a similar way to CV teleportation \cite{braunstein2005quantum}, the procedure consists of two beam splitter operations followed by a sequence of measurements and conditional operations \cite{Braunstein98CV}. As with the CV case, the role of the first beam splitter is to create a shared entangled state between Alice and Bob. This is achieved by applying a beam splitter to a cat state. The measurements consist of parity measurements on the output modes of the second beam splitter, and a further dispersive measurement is made following the parity measurement. Classically transmitting the results of the measurements to Bob, correction operations can be applied to complete the teleportation operation.  We now describe step by step how this procedure achieves quantum teleportation.

\subsection{State Initialization}

The state on mode $ a $ to be teleported is the state
\begin{align}
| \psi_0 \rangle = \mu | 0_L \rangle_\alpha + \nu | 1_L \rangle_\alpha ,
\label{initialcatstate}
\end{align}
where $ \mu, \nu $ are arbitrary complex coefficients satisfying $ |\mu|^2 + |\nu|^2 = 1 $. The states  $|0_{L} \rangle_{\alpha}$ and $|1_{L} \rangle_{\alpha}$ are logical qubit states,  defined as
\begin{align}	
	|0_{L} \rangle_{\alpha} & = \frac{1}{\sqrt{2(1+ e^{-2 |\alpha|^2})}} (|\alpha \rangle + |-\alpha \rangle) \nonumber \\
 & = 
\frac{1}{ \sqrt{ \cosh |\alpha|^2 }}  \sum_{n\in \text{even} } \frac{\alpha^n}{\sqrt{n!}} | n \rangle ,  \nonumber \\
 	|1_{L} \rangle_{\alpha} & = \frac{1}{\sqrt{2(1 - e^{-2 |\alpha|^2})}} (|\alpha \rangle - |-\alpha \rangle) \nonumber \\
   & =  \frac{1}{ \sqrt{ \sinh |\alpha|^2 }} \sum_{n\in \text{odd} } \frac{\alpha^n}{\sqrt{n!}} | n \rangle ,
  \label{evenoddlogical}
\end{align}
where $| \pm \alpha \rangle$ are coherent states. Throughout this paper we consider $ \alpha $ to be a real parameter ($ \alpha \in \mathbb{R} $).  Eq. (\ref{evenoddlogical}) are even and odd cat states as they respectively consist only of Fock states with either even or odd parity. As such they satisfy the orthogonality relation
\begin{align}
\langle \sigma_{L} |_\alpha |  \sigma'_{L} \rangle_\alpha   = \delta_{\sigma \sigma' }. 
\end{align}
We note here that the logical states are eigenstates of the parity measurement operators
\begin{align}
P_+ | 0_L \rangle_\alpha & = | 0_L \rangle_\alpha \nonumber \\
P_- | 1_L \rangle_\alpha & = | 1_L \rangle_\alpha \nonumber \\
P_+ | 1_L \rangle_\alpha & = P_- | 0_L \rangle_\alpha = 0  ,
\label{parityeigenstates}
\end{align}
as they are states of definite parity.  

The relations (\ref{evenoddlogical}) can be inverted to write coherent states in terms of the parity states
\begin{align}
| \pm \alpha \rangle & = \sqrt{\frac{(1+e^{-2|\alpha|^2})}{2}} | 0_L \rangle_\alpha \pm  \sqrt{\frac{(1 -e^{-2|\alpha|^2 }) }{2}} | 1_L \rangle_\alpha .  
\end{align}
For $ | \alpha | $ sufficiently large, to a good approximation, the factors of $ 1 \pm e^{-2 |\alpha|^2} \approx 1 $.   For example, for $ |\alpha | = 2 $, the corrections are at the level of 0.01\%.  Henceforth, we assume this regime and use the approximate relation
\begin{align}
| \pm \alpha \rangle & \approx \frac{1}{\sqrt{2} } ( | 0_L \rangle_\alpha \pm  | 1_L \rangle_\alpha  ).
\label{approxrelation}
\end{align}
The initial state then can be written in terms of the coherent states as 
\begin{align}
| \psi_0 \rangle  = \frac{\mu + \nu}{\sqrt{2}} | \alpha \rangle + 
\frac{\mu - \nu}{\sqrt{2}} | -\alpha \rangle . 
\end{align}

On mode $ b$, initially the cat state 
\begin{align}
\frac{1}{\sqrt{2}} ( | \sqrt{2} \beta \rangle + | -\sqrt{2} \beta \rangle ) 
\end{align}
is prepared, while on mode $ c $ the vacuum state $ | 0 \rangle $ is prepared. We will consider $ \beta $ to be either purely real ($ \beta \in \mathbb{R} $) or imaginary ($ \beta \in \mathbb{I} $).  Slight modifications of the protocol will be needed for each case, which we consider separately.

\subsection{Generation of a shared entangled state}

The first operation in the protocol is to perform a beam splitting operation between modes $ b $ and $ c $.  The mode transformations are the same as given in (\ref{beamsplitter}) but for modes $ b $ and $ c $:
\begin{align}	
    b \rightarrow \frac{1}{\sqrt{2}} (b+c), \nonumber\\
    c\rightarrow \frac{1}{\sqrt{2}} (b-c).
\end{align}
This yields the Bell state 
\begin{align}
&\frac{1}{\sqrt{2}} ( | \sqrt{2} \beta \rangle + | -\sqrt{2} \beta \rangle ) |0\rangle \label{initialbcstate} \\
= & \frac{1}{\sqrt{2}}\left(e^{-|\beta|^2} e^{\sqrt{2}\beta b^\dagger} + e^{-|\beta|^2} e^{-\sqrt{2}\beta b^\dagger}\right) |0\rangle \nonumber\\
\rightarrow &\frac{1}{\sqrt{2}} e^{-|\beta|^2/2} e^{-|\beta|^2/2}\left(e^{\beta(b^\dagger+c^\dagger)}+e^{-\beta(b^\dagger+c^\dagger)}\right)|0\rangle \nonumber\\
=&\frac{1}{\sqrt{2}} (|\beta \rangle |\beta \rangle + |-\beta \rangle |-\beta \rangle)    \nonumber\\
\approx & \frac{1}{\sqrt{2}} (|0_{L}\rangle_{\beta}^{b} |0_{L}\rangle_{\beta}^{c} + |1_{L}\rangle_{\beta}^{b} |1_{L}\rangle_{\beta}^{c}), \label{eq:sharedcat}
\end{align}
where we used (\ref{approxrelation}).  We see that an entangled state is created between Alice and Bob's modes. Superscripts in Eq.~\eqref{eq:sharedcat} indicate the mode for the state. Note that this state can also be prepared through conditional displacement operations \cite{Wang2016jointparity}.

\subsection{Second beam splitter}

 Next we apply another beam splitter operation to create correlation between modes $a$ and $b$, as given in (\ref{beamsplitter}). The output state following the second beam splitter is then written as
\begin{align}	
&  \frac{1}{\sqrt{2}} (  (\mu  |0_{L} \rangle_{\alpha}^{a}  + \nu |1_{L} \rangle_{\alpha}^{a}) ( |0_{L}\rangle_{\beta}^{b} |0_{L}\rangle_{\beta}^{c} + |1_{L}\rangle_{\beta}^{b} |1_{L}\rangle_{\beta}^{c} )\nonumber \\
&   \rightarrow   |\Phi\rangle  =  \frac{1}{2 \sqrt{2} } \Big[ \nonumber \\
& (|0_{L} \rangle_{\chi}^{a} |0_{L} \rangle_{\bar{\chi}}^{b} + |1_{L} \rangle_{\chi}^{a} |1_{L} \rangle_{\bar{\chi}}^{b})  (\mu  |0_{L} \rangle_{\beta}^{c} + \nu |1_{L} \rangle_{\beta}^{c}) \nonumber\\
    + &  (|0_{L} \rangle_{\bar{\chi}}^{a} |0_{L} \rangle_{\chi}^{b} + |1_{L} \rangle_{\bar{\chi}}^{a} |1_{L} \rangle_{\chi}^{b}) (\mu  |0_{L} \rangle_{\beta}^{c} - \nu |1_{L} \rangle_{\beta}^{c}) \nonumber\\
    + & (|0_{L} \rangle_{\chi}^{a} |1_{L} \rangle_{\bar{\chi}}^{b} + |1_{L} \rangle_{\chi}^{a} |0_{L} \rangle_{\bar{\chi}}^{b})  (\mu  |1_{L} \rangle_{\beta}^{c} + \nu |0_{L} \rangle_{\beta}^{c}) \nonumber\\
    - &  (|0_{L} \rangle_{\bar{\chi}}^{a} |1_{L} \rangle_{\chi}^{b} + |1_{L} \rangle_{\bar{\chi}}^{a} |0_{L} \rangle_{\chi}^{b})  (\mu  |1_{L} \rangle_{\beta}^{c} - \nu |0_{L} \rangle_{\beta}^{c}) \Big] , 
    \label{preteleportedstate}
\end{align}
where 
\begin{align}
\chi & = \frac{\alpha + \beta}{\sqrt{2}} \nonumber \\
\bar{\chi} & = \frac{\alpha - \beta}{\sqrt{2}} . 
\label{chidefs}
\end{align}
The transformed state (\ref{preteleportedstate}) is reminiscent of the state prior to measurement in standard qubit teleportation \cite{nielsen2010quantum}. The state consists of a Bell state on Alice's modes and variations of the initial state on Bob's side. 
A measurement of the first two modes collapses the state to one of the four terms in the superposition, leaving the remaining mode in one of four states that correspond to a bit or phase flipped version of Alice's original state.  The task is now to perform a measurement of the first two modes in such a way to induce this collapse using only parity and dispersive measurements.

\subsection{Measurement}
\label{sec:measurement}

The next step involves the first of two measurements that are performed in the teleportation protocol.  We consider two variants of the teleportation protocol, using either single-mode parity measurements (variant 1) or joint parity measurements (variant 2).  We discuss each of these in the following subsections.

\subsubsection{Variant 1: Single-mode parity measurements}
\label{sec:singleparity}

After the second beam splitter operation, we perform single-mode parity measurements on modes $a$ and $b$.  We calculate this by applying the parity projection operators $ P_{\sigma_a}$ and $P_{\sigma_b} $ on the state (\ref{preteleportedstate}), where $ \sigma_a, \sigma_b \in \pm 1 $ label the measurement outcomes.   The four possible parity measurements give the resultant states
\begin{widetext}
\begin{align}	
& P_{\sigma_a} P_{\sigma_b} | \Phi \rangle =   \nonumber \\
& | \psi_{\sigma_a \sigma_b} \rangle = 
\left\{
\begin{array}{cc}
\frac{1}{2 \sqrt{2}} \Big[ |0_{L}\rangle_{\chi} |0_{L}\rangle_{\bar{\chi}} (\mu |0_{L} \rangle_{\beta} + \nu |1_{L} \rangle_{\beta}) + |0_{L}\rangle_{\bar{\chi}} |0_{L}\rangle_{\chi} (\mu |0_{L} \rangle_{\beta} - \nu |1_{L} \rangle_{\beta})\Big]  & \text{if }  \sigma_a = \sigma_b = +  \\
\frac{1}{2 \sqrt{2}} \Big[  |1_{L}\rangle_{\chi} |1_{L}\rangle_{\bar{\chi}} (\mu |0_{L} \rangle_{\beta} + \nu |1_{L} \rangle_{\beta}) + |1_{L}\rangle_{\bar{\chi}} |1_{L}\rangle_{\chi} (\mu |0_{L} \rangle_{\beta} - \nu |1_{L} \rangle_{\beta}) \Big]  &  \text{if } \sigma_a = \sigma_b = -  \\
\frac{1}{2 \sqrt{2}} \Big[  |0_{L}\rangle_{\chi} |1_{L}\rangle_{\bar{\chi}} (\mu |1_{L} \rangle_{\beta} + \nu |0_{L} \rangle_{\beta}) + |0_{L}\rangle_{\bar{\chi}} |1_{L}\rangle_{\chi} (-\mu |1_{L} \rangle_{\beta} + \nu |0_{L} \rangle_{\beta}) \Big]  &  \text{if } \sigma_a = +,  \sigma_b = -  \\
 \frac{1}{2 \sqrt{2}} \Big[   |1_{L}\rangle_{\chi} |0_{L}\rangle_{\bar{\chi}} (\mu |1_{L} \rangle_{\beta} + \nu |0_{L} \rangle_{\beta}) + |1_{L}\rangle_{\bar{\chi}} |0_{L}\rangle_{\chi} (-\mu |1_{L} \rangle_{\beta} + \nu |0_{L} \rangle_{\beta}) \Big]  &  \text{if } \sigma_a = -,  \sigma_b = +  \\
\end{array}
\right. .
\label{paritymeasstate}
\end{align}
\end{widetext}
The probability of each of these four outcomes occurs with probability 
\begin{align}
   p_{\sigma_a \sigma_b} = \langle \Phi |  P_{\sigma_a} P_{\sigma_b} | \Phi \rangle =  \langle  \psi_{\sigma_a \sigma_b} | \psi_{\sigma_a \sigma_b} \rangle \approx \frac{1}{4}, 
\end{align}
which is valid as long as $ |\alpha |, |\beta| $ is taken sufficiently large such that (\ref{approxrelation}) holds.  We observe that only partial collapse of the state (\ref{preteleportedstate}) has occurred, since single-mode parity measurements only can distinguish between two of the four states.  Specifically, the remaining state is an entangled state involving the labels $\chi$ and $\bar{\chi}$ on modes $ a, b $.  The first term in each of the states list in (\ref{paritymeasstate}) has mode $ a $ as a cat state in terms of a coherent state amplitude $ \chi $, while for the second term the cat state has an amplitude $ \bar{\chi} $.  Note that the superscripts denoting the modes have been removed here and in the following sections to simplify the notation. The ordering of the states is taken to be $ a,b,c $ when the full system wavefunction is written.

\begin{figure}[th!]
    \centering
    \includegraphics[width=\columnwidth]{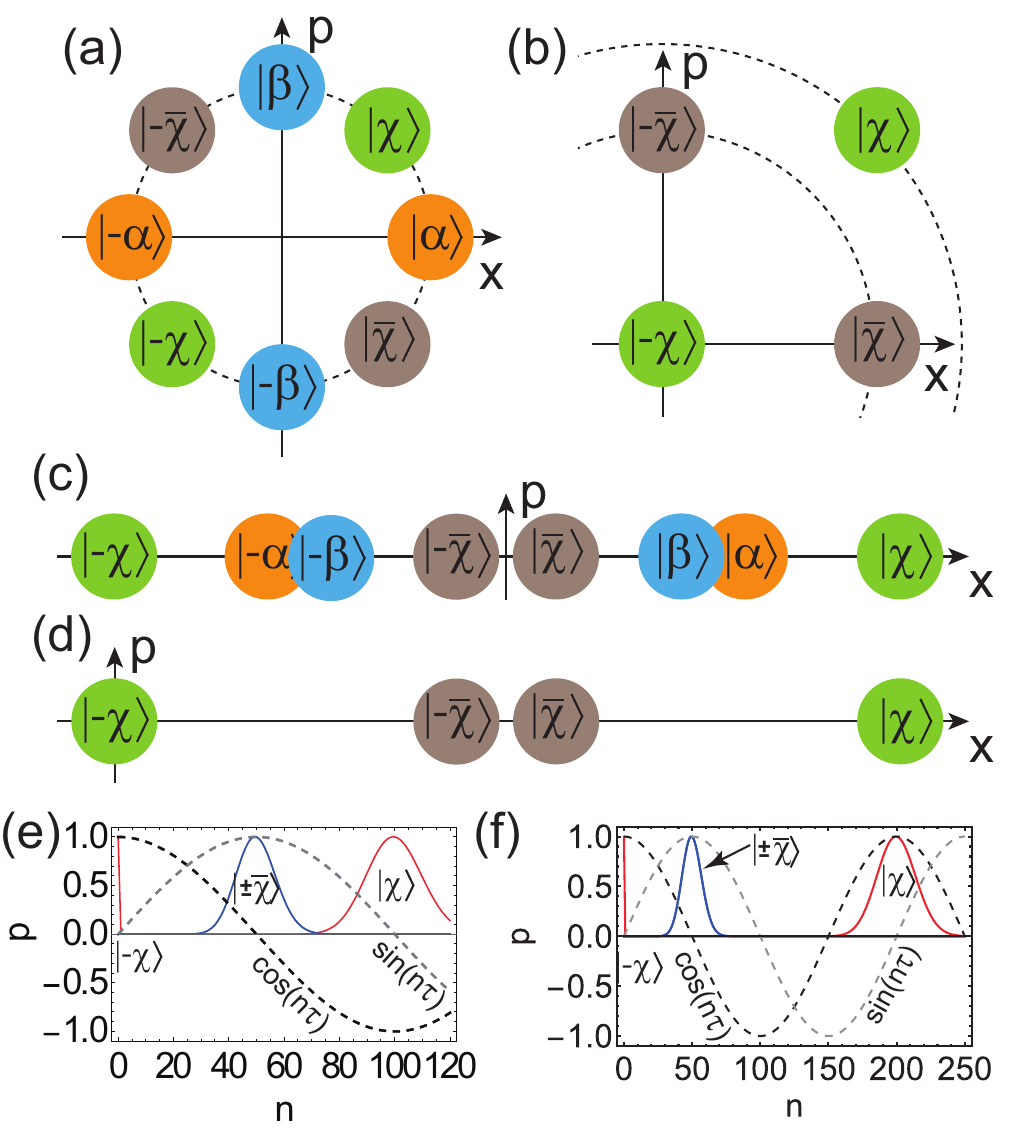}
\caption{The various coherent states that appear in our quantum teleportation protocol. \textbf{(a)}\textbf{(c)}  The coherent states $ | \pm \alpha \rangle $ form the original logical states on mode $ a $, where we take $ \alpha $ to be real.  The coherent states $ | \pm \beta \rangle $ form the Bell state on modes $ b$ and $ c$ and can be taken to be either \textbf{(a)} purely imaginary or \textbf{(c)} purely real.  The combinations of these $ | \pm \chi \rangle $ and $ | \pm \bar{\chi} \rangle $ as defined in (\ref{chidefs}) resulting from the second beam splitter is also shown in each case.  \textbf{(b)}\textbf{(d)} The amplitudes of the $ | \pm \chi \rangle $ and $ | \pm \bar{\chi} \rangle $ states after performing a displacement $ D(\chi) $. In  \textbf{(b)}, the inner dashed line shows the amplitude of the states $ | \bar{\chi}|  = \sqrt{2} \alpha $.  The outer dashed line is the amplitude of the state $ |  \chi \rangle $ which has a radius $ 2 \alpha $. \textbf{(e)}\textbf{(f)}The probability distributions in the Fock basis $ p_n = | \langle n | \xi \rangle |^2 $ for the coherent states with $ \xi = 0 , 2 \chi, \chi \pm \bar{\chi} $. Parameters chosen are \textbf{(e)} $ \alpha = 5, \beta = 5i  $; \textbf{(f)} $ \alpha = 5, \beta = 5 $. }
    \label{fig:coherent}
\end{figure}

A plot of the coherent states that appear in our protocol are shown in Fig. \ref{fig:coherent}.  We see that for the choice of $ \alpha > 0  $ real and purely imaginary $ \beta = i \alpha $, the coherent states $ | \pm \chi \rangle $ and $ | \pm \bar{\chi} \rangle $ lie in distinct parts of phase space to each other (Fig. \ref{fig:coherent}(a)).  Alternatively, for the choice of $ \alpha \sim \beta $ both real,  $ | \pm \chi \rangle $ appear in distinct parts of phase space, but $ | \pm \bar{\chi} \rangle $ tends to be close together.  If we are able to perform a measurement identifying whether mode $ a $ has amplitude $\pm \chi$ or  $\pm \bar{\chi}$, we should be able to collapse the state (\ref{paritymeasstate}) further and obtain the teleported state. 

Such a measurement can be accomplished by performing a displacement $ D(\chi) $ followed by a measurement in the Fock basis.  Fig.  \ref{fig:coherent}(b)(d) shows the amplitude of the  $| \pm \chi\rangle $ or  $| \pm \bar{\chi}\rangle $ states after the displacement.  For the choice $ \beta = i \alpha $ (Fig.  \ref{fig:coherent}(b)), we see that the $| \pm \chi\rangle $ states now have amplitudes of $ 0 $ and $ 2 \alpha$ respectively, while the $| \pm \bar{\chi}\rangle $ states both have an amplitude of $ \sqrt{2} \alpha $. As these have different amplitudes, an estimate of the Fock number will collapse the state to one of the coherent states. For $ \beta = \alpha $ taking real values, the $| \pm \chi\rangle $ states now have amplitudes of $0$ and $ 2 \sqrt{2} \alpha$, while the $| \pm \bar{\chi}\rangle $ states have an amplitude of $ \sqrt{2} \alpha $.

The number distribution of the states $ | \pm \chi  \rangle, | \pm \bar{\chi} \rangle $ after a displacement of $ D(\chi) $ is shown in Figs. \ref{fig:coherent}(e)(f).  For the $ \beta = i \alpha  $ case (Fig. \ref{fig:coherent}(e)), we see that there is a sharp peak at $ n = 0 $ for the vacuum state, as well as peaks at $ n = 2 \alpha^2 $ and $ n = 4 \alpha^2$ for the states $ | \pm \bar{\chi} \rangle $ and $ |+ \chi \rangle $ respectively. Therefore, taking as an example the state with $\sigma_a=\sigma_b=+$ in Eq.~\eqref{paritymeasstate}, measuring a Fock state $| n \rangle$ with $n\approx 2\alpha^2$ will collapse the state in modes $c$ to $ \mu |0_{L} \rangle_{\beta} - \nu |1_{L} \rangle_{\beta} $, while $n=0$ or $n\approx 4 \alpha^2$ will collapse the state in modes $c$ to $ \mu |0_{L} \rangle_{\beta} + \nu |1_{L} \rangle_{\beta} $.  For the $ \beta = \alpha $ case, we see that there is a sharp peak at $ n = 0 $ for the vacuum state  (Fig. \ref{fig:coherent}(f)), as well as peaks at $ n = 2 \alpha^2 $ and $ n = 8 \alpha^2$ for the states $ | \pm \bar{\chi} \rangle $ and $ | + \chi \rangle $ respectively.  Again, a measurement in the Fock basis will distinguish the coherent state amplitudes in this case.

\begin{table}[t!]
\begin{center}
\begin{tabular}{ |c|c|c|c|c| } 
\hline
mode $ c $ state & $\sigma_a $  & $\sigma_b $ & $ \sigma_a' $ & correction \\
\hline
\multirow{2}{*}{$\mu |0_L \rangle_\beta + \nu | 1_L \rangle_\beta $} & $+$ & $+$  & $+$ & \multirow{2}{*}{$I_L $} \\ 
& $-$ & $-$ & + & \\  
\hline
\multirow{2}{*}{$\mu |0_L \rangle_\beta - \nu | 1_L \rangle_\beta $} & $+$ & $+$  & $-$ & \multirow{2}{*}{$Z_L $} \\ 
& $-$ & $-$ & $-$ & \\  
\hline
\multirow{2}{*}{$\mu |1_L \rangle_\beta + \nu | 0_L \rangle_\beta $} & $+$ & $-$ & $+$ & \multirow{2}{*}{$X_L $} \\ 
& $-$ & $+$ & $+$ & \\  
\hline
\multirow{2}{*}{$ - \mu |1_L \rangle_\beta + \nu | 0_L \rangle_\beta $} & $+$ & $-$  & $-$ & \multirow{2}{*}{$X_L Z_L  $} \\ 
& $-$ & $+$ & $-$ & \\  
\hline
\end{tabular}
\end{center}
\caption{The approximate state on mode $ c $ using Variant 1 of the teleportation protocol, involving two single-mode parity measurements. We omit the states on modes $ a $ and $ b $ which are approximately disentangled after the measurement.  $ \sigma_a $ and $ \sigma_b $ are the outcomes of the parity measurement on modes $ a $ and $ b$.  $ \sigma_a' $ is the outcome of the dispersive measurement on mode $ a $.   }
\label{tab:finalstates}
\end{table}

While a measurement in the Fock basis would cause the desired collapse, this type of measurement is typically not easily performed for the system that we consider.  For this reason, we now provide an alternative method based on single-shot dispersive measurements. We now show that a single-shot dispersive measurement can be performed to achieve the desired collapse. Consider the measurement (\ref{dispmeas}) with interaction time chosen as 
\begin{align}
    \tau = \frac{\pi}{2 \bar{n} } , 
    \label{optimaltime}
\end{align}
where
\begin{align}
\bar{n} & =\frac{1}{2} ( | \chi + \bar{\chi} |^2 +| \chi - \bar{\chi} |^2 )  \nonumber \\
& = | \alpha |^2 +  | \beta |^2 
\end{align}
is the average boson number of the states $ |\chi \pm \bar{\chi} \rangle $.  

For example, with the choice $ \beta = i \alpha $, the optimal time is $ \tau = \pi/(4 \alpha^2) $.  This choice puts the zero of the cosine function for the $ M_+ $ outcome coinciding with the $ | \pm \bar{\chi } \rangle $ states (Fig. \ref{fig:coherent}(e)).  Meanwhile, the sine function is a zero at location of the $ | \pm \chi \rangle $ states. Hence, we approximately have for  $ \beta = i \alpha $:
\begin{align}
M_+ D(\chi) | \sigma_L \rangle_\chi & \approx - D(\chi)| (1-\sigma)_L \rangle_\chi \nonumber \\
M_+ D(\chi) | \sigma_L \rangle_{\bar{\chi}} & \approx 0   \nonumber \\
M_- D(\chi) | \sigma_L \rangle_\chi & \approx 0 \nonumber \\
M_- D(\chi) | \sigma_L \rangle_{\bar{\chi}} & \approx D(\chi)  | \sigma_L \rangle_{\bar{\chi}} ,
\label{measurementdisp}
\end{align}
with $ \sigma \in \{0,1 \} $. 

Meanwhile, for $ \beta = \alpha $, the optimal time is again $ \tau = \pi/(4 \alpha^2) $.  As can be seen from Fig. \ref{fig:coherent}(f), the cosine function again has a zero at the location of the $ | \pm \bar{\chi } \rangle $ states.  The sine function again takes zero values at the location of the $ | \pm \chi \rangle $ peaks.  In this case, we approximately have for $ \beta = \alpha$:
\begin{align}
M_+ D(\chi) | \sigma_L \rangle_\chi & \approx  D(\chi)| \sigma_L \rangle_\chi \nonumber \\
M_+ D(\chi) | \sigma_L \rangle_{\bar{\chi}} & \approx 0   \nonumber \\
M_- D(\chi) | \sigma_L \rangle_\chi & \approx 0 \nonumber \\
M_- D(\chi) | \sigma_L \rangle_{\bar{\chi}} & \approx D(\chi)  | \sigma_L \rangle_{\bar{\chi}} .
\label{measurementdispreal}
\end{align}
Hence, for both $ \beta = i \alpha, \alpha$ the displaced measurements are effective at distinguishing between the $ | \pm \chi\rangle  $ and $ | \pm \bar{\chi} \rangle  $ states.   

Performing such a measurement on mode $ a $, the resulting states together with the measurement outcomes are listed in Table \ref{tab:finalstates}. The state on mode $ c $, up to correction operations, is the original initial state on Alice's mode $ a $.  
Note that since the measurement is performed on mode $ a $, which is traced out, the phase factor $e^{-in\omega t}$ does not affect the teleported state, which is on mode $ c $.

The collapse relations (\ref{measurementdisp}) and (\ref{measurementdispreal}) are only approximate as the zeros of the measurement function only coincide with the centers of the peaks, and there is some residual entanglement after projecting the state (\ref{paritymeasstate}).  This can be improved by repeating the dispersive measurements
\begin{align}	
    M_{k_+,k_-} &= \sqrt{{N \choose k_+}} M_{+}^{k_{+}} M_{-}^{k_{-}}  \nonumber\\
    & = \sum_{n} \sqrt{{N \choose k_+}} \text{cos}^{k_+}(\tau n) \text{sin}^{k_-}(\tau n) |n\rangle \langle n|,
    \label{multiplemeasdef}
\end{align}
where $k_\pm$ are the number of times the outcome $ M_\pm$ is obtained, respectively, and the total number of measurements is $ N = k_+ + k_- $.  The binomial factor is present because there can be more than one outcome that have the same $k_\pm$.  For example, for $ N = 2$ measurements, $ M_+ M_-$ and $ M_- M_+$ are both outcomes with $ k_+ = k_- = 1$.  The binomial factor ensures that $ \sum_{k_+=0}^N M_{k_+,N-k_+}^\dagger M_{k_+,N-k_+} = I $.  
The effect of multiple measurements is to cause a further collapse on the Fock basis.  The trigonometric factor in (\ref{multiplemeasdef}) can be approximated by a Gaussian \cite{PhysRevA.94.013617,kondappan2023imaginary}
\begin{align}
 \sqrt{{ N\choose k_+}}  \text{cos}^{k_+}(x) \text{sin}^{k_-}(x) \propto \exp( -N (x-x_{k_+ k_-})^2 ) , 
\end{align}
where the center of the Gaussian is 
\begin{align}
x_{k_+ k_-} = \frac{1}{2} \arccos \left( \frac{k_+ - k_-}{N} \right)  . 
\label{gaussianpeak}
\end{align}
Then one can find which Fock state the collapse occured at by the relation $ n = x_{k_+ k_-} /\tau $.  From this information, one may determine which of the coherent states $ | \pm \chi \rangle ,  | \pm \bar{\chi} \rangle $ the collapse occurred at.

\subsubsection{Variant 2: Joint parity measurements}
\label{sec:joint}

As can be seen in Table \ref{tab:finalstates}, there are a total of 8 measurement outcomes, according to the three binary outcomes for $ \sigma_a,  \sigma_b, \sigma_a' $.  This is in contrast to standard qubit teleportation where a measurement of Alice's qubits gives four measurement outcomes.  This suggests that our measurement scheme is suboptimal since three binary-outcome measurements are being made.  Here we show that it is possible to reduce the number of binary measurements to two.  

On closer inspection of Table \ref{tab:finalstates}, examining the outcomes for $ \sigma_a,  \sigma_b $, one may see that there is indeed a redundancy that can be eliminated.  We see that for the four mode $ c $ states, the product $ \sigma_a \sigma_b $ yields the same value.  This suggests that by performing a joint parity measurement of $ \sigma_{ab} = \sigma_a \sigma_b $, without measuring $ \sigma_a, \sigma_b $ individually, one can achieve the same result as Table \ref{tab:finalstates}.  Such joint parity measurements have also been demonstrated experimentally \cite{Wang2016jointparity,Huai2024jointparity}. Such a method can be calculated by applying the joint parity projection operator $ P_{\sigma_{ab}}  $ on the state (\ref{preteleportedstate}). 

After applying a joint parity measurement on modes $a$ and $b$, the state (\ref{preteleportedstate}) will collapse onto two possible states with joint parity $\sigma_{ab} \in \pm $. The resulting state is 
\begin{widetext}
\begin{align}	
 P_{\sigma_{ab}}|\Phi\rangle  = |\psi_{\sigma_{ab}} \rangle = 
\left\{
\begin{array}{cc}
\frac{1}{2 \sqrt{2}} \Big[  (|0_{L}\rangle_{\chi} |0_{L}\rangle_{\bar{\chi}} + |1_{L}\rangle_{\chi} |1_{L}\rangle_{\bar{\chi}})  (\mu |0_{L} \rangle_{\beta} + \nu |1_{L} \rangle_{\beta})  & \nonumber \\
 + (|0_{L}\rangle_{\bar{\chi}} |0_{L}\rangle_{\chi} +|1_{L}\rangle_{\bar{\chi}} |1_{L}\rangle_{\chi}) (\mu |0_{L} \rangle_{\beta} - \nu |1_{L} \rangle_{\beta})  \Big]  &  \quad\text{if } \sigma_{ab} =  +  \\
\frac{1}{2 \sqrt{2}} \Big[  ( |0_{L}\rangle_{\chi} |1_{L}\rangle_{\bar{\chi}} + |1_{L}\rangle_{\chi} |0_{L}\rangle_{\bar{\chi}} ) (\mu |1_{L} \rangle_{\beta} + \nu |0_{L} \rangle_{\beta})  & \nonumber \\ 
+ (|0_{L}\rangle_{\bar{\chi}} |1_{L}\rangle_{\chi} +|1_{L}\rangle_{\bar{\chi}} |0_{L}\rangle_{\chi} ) (-\mu |1_{L} \rangle_{\beta} + \nu |0_{L} \rangle_{\beta}) \Big]  &  \quad\text{if } \sigma_{ab}=-  \\
\end{array}
\right. . \\
\label{jointparitymeasstate}
\end{align}
\end{widetext}
The probability of these two outcomes are
\begin{equation}
   p_{\sigma_{ab}} = \langle \Phi |  P_{\sigma_{ab}} | \Phi \rangle = \langle \psi_{\sigma_{ab}}  |\psi_{\sigma_{ab}} \rangle \approx  \frac{1}{2} .
\end{equation}
which holds as long as $ |\alpha |, |\beta| $ is taken sufficiently large.

Similarly to (\ref{paritymeasstate}), this involves a superposition of two
versions of the initial state. The difference here is that modes $ a $ and $ b $ are left in one of the Bell states. The superposition of these terms can be further collapsed by distinguishing between the amplitudes $ \chi $ and $ \bar{\chi} $, using a procedure similar to that described in Sec. \ref{sec:singleparity}. In Table \ref{tab:finalstatesjoint}, we summarize the various outcomes for the joint parity measurement and the displaced measurement.

\begin{table}[t!]
\centering
\begin{tabular}{ |c|c|c|c| }
\hline
mode $ c $ state & $\sigma_{ab} $  & $ \sigma_a' $ & correction \\
\hline
\multirow{1}{*}{$\mu |0_L \rangle_\beta + \nu | 1_L \rangle_\beta $} & $+$  & $+$ & \multirow{1}{*}{$I_L $} \\ 
\hline
\multirow{1}{*}{$\mu |0_L \rangle_\beta - \nu | 1_L \rangle_\beta $} & $+$  & $-$ & \multirow{1}{*}{$Z_L $} \\ 
\hline
\multirow{1}{*}{$\mu |1_L \rangle_\beta + \nu | 0_L \rangle_\beta $} & $-$  & $+$ & \multirow{1}{*}{$X_L $} \\ 
\hline
\multirow{1}{*}{$ - \mu |1_L \rangle_\beta + \nu | 0_L \rangle_\beta $} & $-$  & $-$ & \multirow{1}{*}{$X_L Z_L  $} \\  
\hline
\end{tabular}
\caption{The approximate state on mode $ c $ using Variant 2 of the teleportation protocol, involving the joint parity measurement. We omit the states on modes $ a $ and $ b $ which are approximately disentangled after the measurement.  $ \sigma_{ab} $ are the outcomes of the joint parity measurement on the first two modes $ a $,$ b$.  $ \sigma_a' $ is the outcome of the dispersive measurement on mode $ a $.   }
\label{tab:finalstatesjoint}
\end{table}

\subsection{Correction operation}
\label{sec:corr}

The final step is to perform a conditional operation to correct the states listed in Table \ref{tab:finalstates} and Table \ref{tab:finalstatesjoint}. At the logical level, it is clear that operations corresponding to Pauli $ Z_L $ and $ X_L $ operators are required.  At the physical level, this can be realized by 
\begin{align}
\label{zlcorr}
    Z_L = e^{-i c^\dagger c \pi }
\end{align}
which achieves $ Z_L | 0_L \rangle_\beta = | 0_L \rangle_\beta $ and $ Z_L | 1_L \rangle_\beta = -| 1_L \rangle_\beta $ according to the definition (\ref{evenoddlogical}). 

For the $\beta \in \mathbb{I} $ case, the $ X_L $ correction can be achieved by noticing that the first relation in (\ref{measurementdisp}) performs a logical spin flip.  We thus have
\begin{align}
 X_L \approx D(-\beta) M_+ (\frac{\pi}{2\bar{n}}) D (\beta)  ,  
\end{align}
where the displacement $ D(-\beta) $ returns the state to the original cat state amplitudes.  As the measurement is probabilistic, ordinarily we would need to consider the $ M_-$ outcome as well.  However, this outcome has a low probability since the sine function has zeros at the locations of the coherent states, \ie $ M_- D(\beta) | \sigma_L \rangle_\beta  \approx 0 $. 

For the $\beta \in \mathbb{R} $ case, from (\ref{measurementdispreal}) we see that there is no operation that performs a logical spin flip.  This is because in Fig. \ref{fig:coherent}(f), the $ | \chi \rangle $ state falls on the positive part of the cosine function.  A spin flip corresponds to a negative phase, which can be achieved by adjusting the cosine function period to double the period.  Hence, choosing $\tau = \pi/4\bar{n}$, we have 
\begin{align}
 X_L \approx D(-\beta) M_+(\frac{\pi}{4\bar{n}}) D (\beta). 
\end{align}
The effect of the measurement is $M_+(\frac{\pi}{4\bar{n}}) D (\beta) | \sigma_L \rangle_\beta  \approx - D(\beta)| (1-\sigma)_L \rangle_\beta$. Therefore, $X_L$ produces a bit flip operation between $| 0_L \rangle_\beta$ and $| 1_L \rangle_\beta$.

\section{Teleportation performance using single-mode parity measurements} \label{singleparityresults}

In this section, we numerically analyze the teleportation protocol as introduced in Sec. \ref{section3}.  Of the two variants of the teleportation protocol involving the single-mode parity measurements (Variant 1, Sec. \ref{sec:singleparity}) and the joint parity measurements (Variant 2, Sec. \ref{sec:joint}), in this section we consider the former.  The latter will be analyzed in Sec. \ref{jointparityresults}.  

We have numerically verified that the approximations starting from the initial states (\ref{initialcatstate}) and (\ref{initialbcstate}) leading to the state (\ref{paritymeasstate}) are extremely accurate, as long as one works within the approximation (\ref{approxrelation}).  For example, the fidelity of the exact state with respect to (\ref{paritymeasstate}) is at the $ 99.8\% $ level choosing $ \alpha = 1, \beta = i$, and improves further as $|\alpha|,|\beta| $ are increased. In most systems, the amplitudes of the cat states are at the few photon level \cite{Vlastakis2013, Sun2014, Kirchmair2013, bild2023schrodinger, Yang2025}, such that the values of $|\alpha|$ and $|\beta|$ that we chose are in the range $[3,4]$.  As can be seen for $|\alpha| = 3$ in Ref. \cite{Yang2025}, the approximate relation (\ref{approxrelation}) performs well experimentally. Most of the approximation in obtaining the teleported states shown in Table \ref{tab:finalstates} arise from the final measurement, to perform the collapse in (\ref{paritymeasstate}).  We therefore start with the expressions (\ref{paritymeasstate}) and then perform numerical computations of the displacement operation and final measurement.  We focus on the  $\sigma_a = \sigma_b = + $ outcome and examine the performance of the teleportation protocol. For the case $ \sigma_a'=+ $ there is no correction necessary such that $ C_{+++} = I_L = I$.  Other measurement outcomes give similar results.

We define the fidelity for the measurement outcome $( \sigma_a, \sigma_b,  \sigma_a') $ as
\begin{align}
F_{\sigma_a \sigma_b \sigma_a'} =  &  ( \mu^* \langle 0_L |_\beta  +  \nu^* \langle 1_L | ) C_{\sigma_a \sigma_b \sigma_a'}   \rho_{ \sigma_a'| \sigma_a \sigma_b}  \nonumber \\
& \times C^\dagger_{\sigma_a \sigma_b \sigma_a'} | ( \mu | 0_L \rangle_\beta  + \nu | 1_L \rangle_\beta ) ,
\label{fiddef}
\end{align}
where 
\begin{align}
 \rho_{ \sigma_a'| \sigma_a \sigma_b}  = \frac{\text{Tr}_{a,b} | \psi_{ \sigma_a' | \sigma_a \sigma_b} \rangle  \langle \psi_{\sigma_a' | \sigma_a \sigma_b  } |}{p_{\sigma_a' | \sigma_a \sigma_b } }
\end{align}
is the state on mode $ c $ after tracing out modes $ a,b $. $ C_{\sigma_a \sigma_b \sigma_a'}  $ is the correction operation that appears in Table \ref{tab:finalstates}, that is physically implemented using the operations given in Sec. \ref{sec:corr}.  The state after applying the displacement and dispersive measurement is
\begin{align}
| \psi_{\sigma_a' | \sigma_a \sigma_b  } \rangle = \frac{M_{\sigma_a'} D(\chi) | \psi_{\sigma_a \sigma_b } \rangle}{\sqrt{p_{\sigma_a \sigma_b}  }} ,
\end{align}
where the state $  | \psi_{\sigma_a \sigma_b } \rangle $ is given in (\ref{paritymeasstate}).  This state has a conditional probability
\begin{align}
p_{\sigma_a' | \sigma_a \sigma_b }=  \langle \psi_{\sigma_a' | \sigma_a \sigma_b  }   | \psi_{\sigma_a' | \sigma_a \sigma_b  } \rangle . 
\label{probdef}
\end{align}

\begin{figure}[t]
    \centering
    \includegraphics[width=0.8\columnwidth]{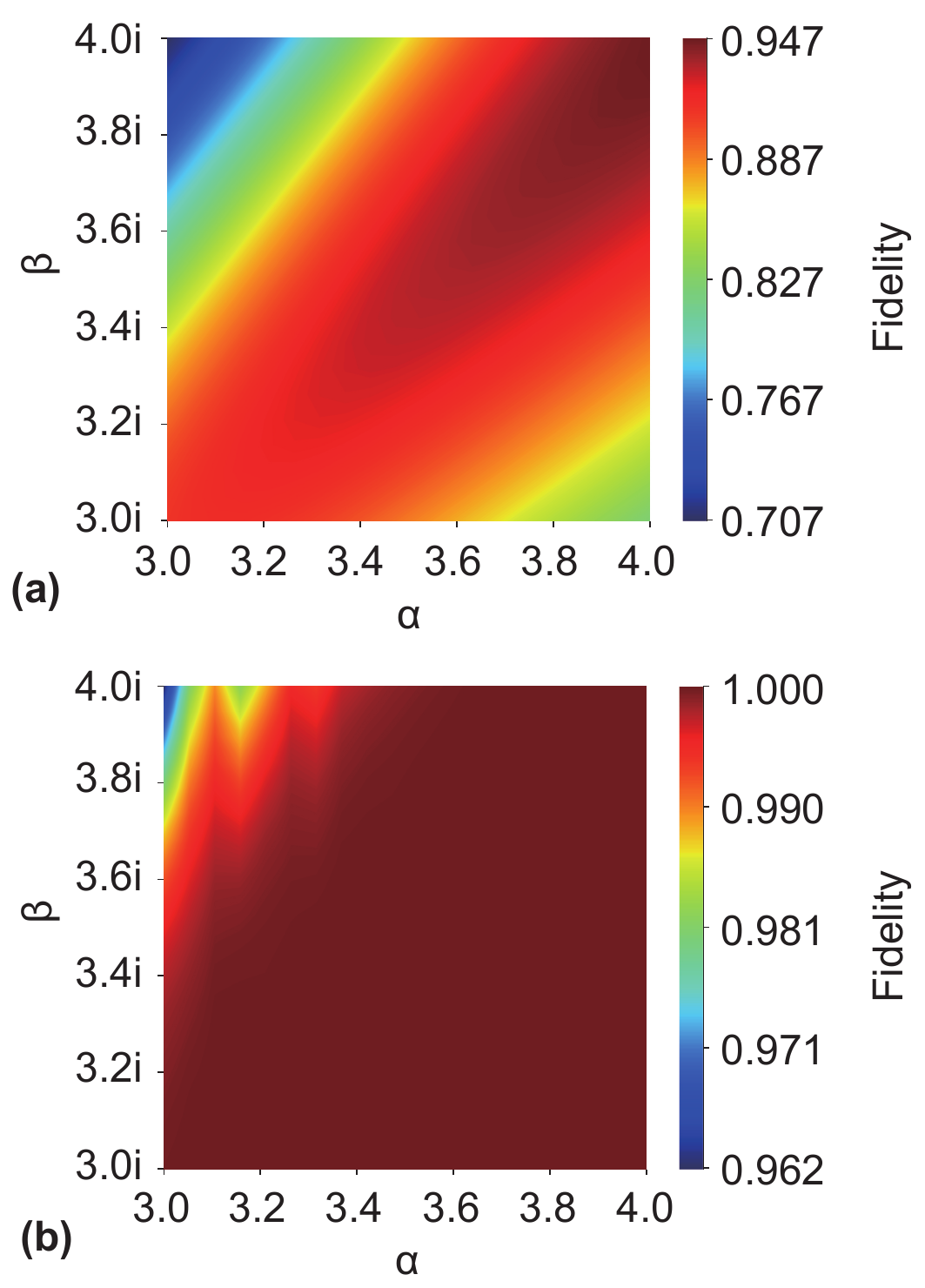}
    \caption{Fidelity (\ref{fiddef}) in the space of  $\alpha \in \mathbb{R} $ and $\beta \in \mathbb{I} $ for Variant 1 teleportation with single-mode parity measurements. Parameters are \textbf{(a)} $N=1, k_+=1, k_-=0 $; \textbf{(b)} $ N=1000, k_+=1000, k_-=0  $.  Common parameters are  $ \mu=1/2, \nu=\sqrt{3}/2, \tau = \pi/(4\alpha^2)$.}
    \label{figure3}
\end{figure}
\begin{figure}[t]
    \centering
    \includegraphics[width=0.8\columnwidth]{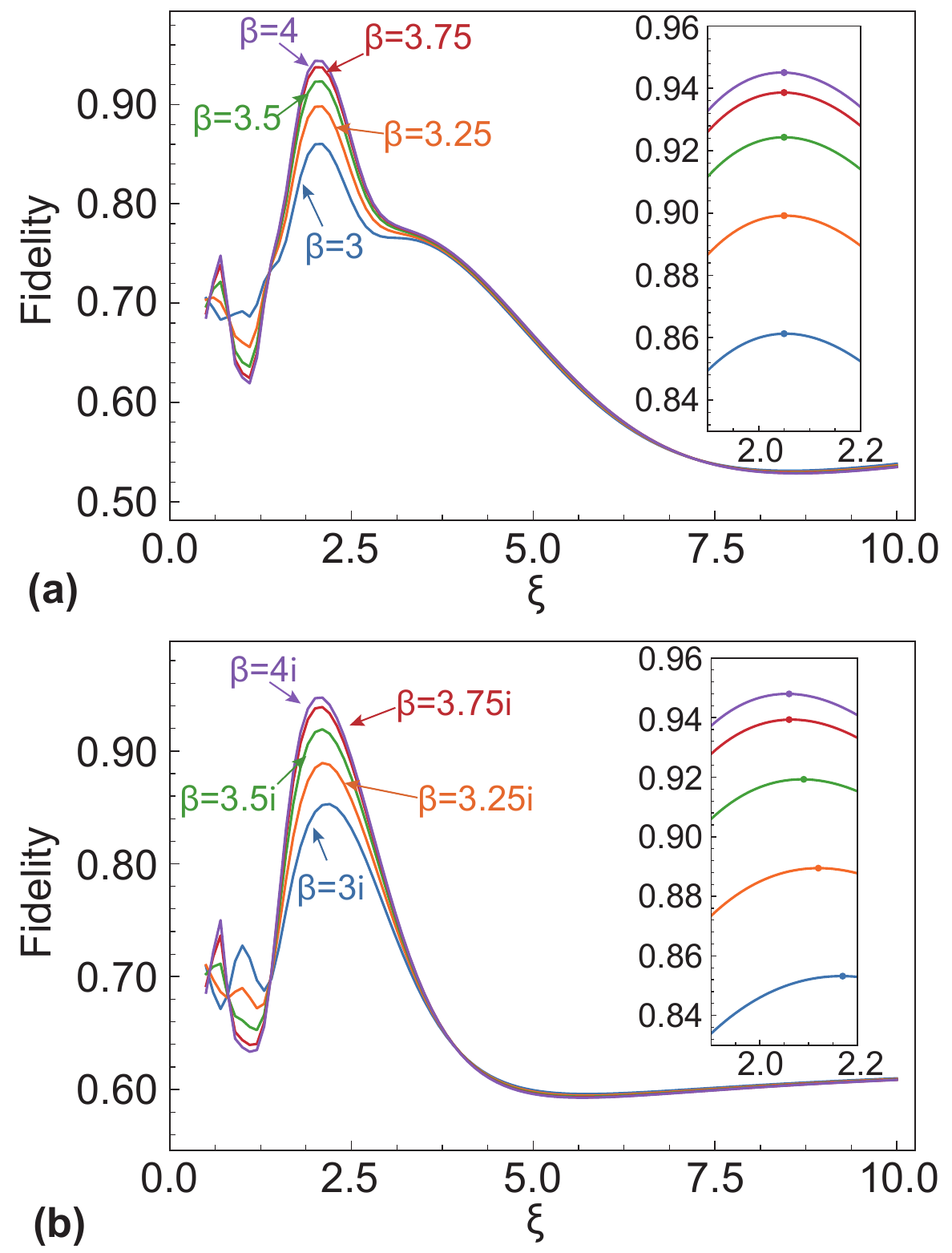}
    \caption{Fidelity (\ref{fiddef}) as a function of $\xi$ for Variant 1 teleportation with single-mode parity measurements. A zoomed-in plot is shown in the inset for better visibility of the optimal points. Parameters are \textbf{(a)} $ \beta \in \mathbb{R} $; \textbf{(b)} $ \beta \in \mathbb{I} $. Common parameters are $ \mu=1/2, \nu=\sqrt{3}/2, \alpha=4, N=1,  k_+=1, k_-=0, \tau = \pi/(\xi \bar{n} ) $. }
        \label{figure5}
\end{figure}

\subsection{$ \alpha \in \mathbb{R} $ and $ \beta \in \mathbb{I}  $, $ N = 1$ measurements }

In Fig. \ref{figure3}(a) we plot the fidelities for various choices of $ \alpha \in \mathbb{R} $ and $ \beta \in \mathbb{I}  $ for the fixed interaction time $ \tau = \pi/(4 \alpha^2) $. We note that this interaction time is not the optimal time as derived in (\ref{optimaltime}) unless $ |\beta| = |\alpha |$. For the final dispersive measurement we consider the measurement outcome $ \sigma_a' = + $ corresponding to a collapse on the $ |\pm \chi \rangle $ states (see Fig. \ref{fig:coherent}(e)).  We see that high fidelities are obtained for various choices of $ \alpha, \beta $.  The best fidelities are obtained in regions with larger $ |\alpha| $ and $ |\beta| $, and it is maximized when $\beta = i\alpha$. Even higher fidelities are expected with further increase of $|\alpha|$ and $|\beta| $.

\subsection{$ \alpha \in \mathbb{R} $ and $ \beta \in \mathbb{I}  $ , $ N = 1000 $ measurements}
\label{sec:jointmanymeas}

Fig. \ref{figure3}(b) shows the results for the same parameters as Fig. \ref{figure3}(a), but the dispersive measurements at the end are repeated $ N = 1000$ times. We again use the fixed interaction time $ \tau = \pi/(4 \alpha^2) $, which is suboptimal as derived in (\ref{optimaltime}) unless $ |\beta| = |\alpha |$.  Here we consider the outcome $ k_+ = N, k_- = 0 $.  This corresponds to the Gaussian peak at the position $ x = \tau n = 0 $ according to (\ref{gaussianpeak}).  This collapses the state on the vacuum state, which is originally the state $ |- \chi \rangle $ before the displacement, and hence suppresses the $ | \pm \bar{\chi} \rangle $ (see Fig. \ref{fig:coherent}(e)).   We see that increasing the number of measurements produces an improvement of the fidelity in comparison to Fig. \ref{figure3}(a).   As can be seen in Fig. \ref{figure3}(b), near unit fidelities are attainable in a larger range of parameters.  The reason for this is that for the interaction time $\tau = \pi/(4\alpha^2)$, the measurement function depends only on the parameter $\alpha$, and the larger $\alpha$ ensures that $M_+$ gives zero on $|\pm \chi \rangle$ states, thereby leading to the maximum suppression of the $|\pm \chi \rangle$ states.

\begin{figure}[t]
    \centering
    \includegraphics[width=0.8\columnwidth]{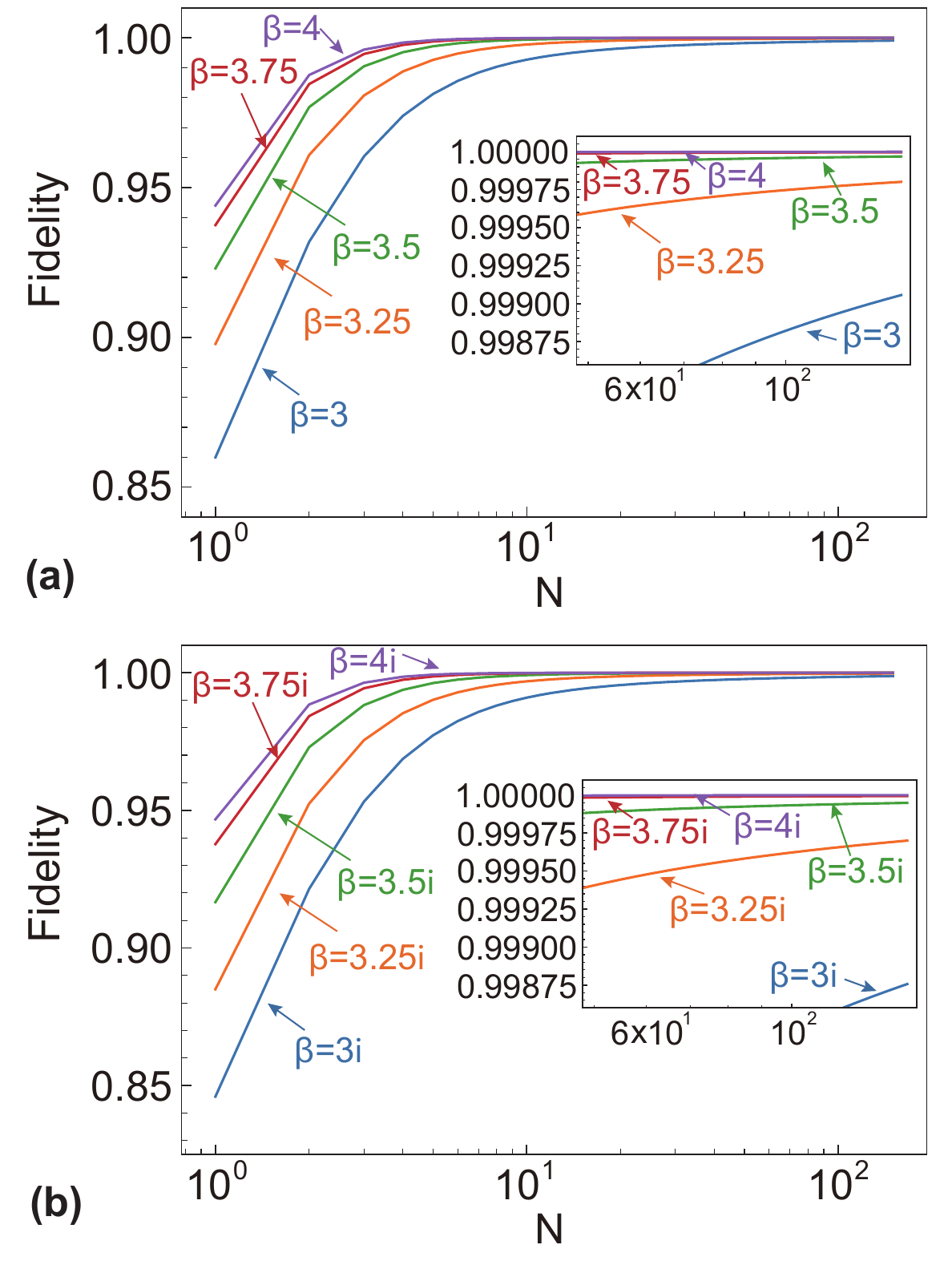}
    \caption{Fidelity (\ref{fiddef}) as a function of $N$ for Variant 1 teleportation with single-mode parity measurements. A zoomed-in plot is shown in the inset for better visibility of the part of the curves that approach 1. The horizontal axis takes a logarithmic scale. Parameters are \textbf{(a)} $ \beta \in \mathbb{R} $; \textbf{(b)} $ \beta \in \mathbb{I} $. Common parameters are $\mu=1/2, \nu=\sqrt{3}/2,\alpha=4, k_+ = N, k_-=0, \tau = \pi/(2\bar{n}) $.}
        \label{figure6}
\end{figure}

\subsection{Dependence on the interaction time $ \tau = \pi/(\xi \bar{n}) $}
\label{sec:taudependence}

To verify the extent to which  (\ref{optimaltime}) can predict the optimal time, we calculate the fidelity for various interaction times. In Fig. \ref{figure5}(a) we plot the fidelity as a function of $\xi$, where $ \tau = \pi/( \xi \bar{n}) $.  We find that the optimal points are at the point $ \xi \approx 2.05 $ for all the chosen $\beta \in \mathbb{R}$, and the best fidelity is at the level of 95\%.  This is close to the formula (\ref{optimaltime}), which predicts $ \xi =2 $. The discrepancy arises due to (\ref{optimaltime}) being based on the average boson number of the $ | \pm \bar{\chi} \rangle $ states.  Small deviations of this may produce slightly better results.  

In Fig. \ref{figure5}(b) we show the fidelities as a function of  $ \xi $ for parameter choices with $\beta \in \mathbb{I}$.  We find that the optimal points are sensitive to the parameter $\xi$ unlike we observed in Fig. \ref{figure5}(a). The reason for this is investigated in more detail in Appendix \ref{appendixA}.  In short, when $\beta \in \mathbb{I}$, the effect of the peak $ | \chi \rangle $ in Fig. \ref{fig:coherent}(e) is closer to the $ | \pm \bar{\chi} \rangle $ peak, and there is a stronger interference effect between the superposition of coherent states, which affects the optimal time.  Nevertheless, $ \tau = \pi/( 2 \bar{n}) $ gives an excellent approximation for the optimal time, with similar fidelities being achieved as the $\beta \in \mathbb{R}$ case.

\subsection{Dependence on the number of measurements $ N $}

We have already see in Sec. \ref{sec:jointmanymeas} that higher fidelities are obtained when a large number of measurements are made in the final step.  Here we examine the dependence with the total number of measurements $N $.  
The fidelity as a function of $N $ is shown in Fig. \ref{figure6}. We choose the nearly optimal interaction time of $ \tau  = \pi/(2\bar{n})$.  We see that the fidelity approaches 1 as the number of measurements is increased, as the measurements eventually collapse the state in the Fock basis and distinguish between the $ | \pm \chi \rangle $ and $ | \pm \bar{\chi} \rangle $ states. As shown in the insets of Fig. \ref{figure6}(a)(b), the larger values $|\beta|$ exhibit a faster convergence rate. 
Comparing the convergence for both real and imaginary $\beta $, we see that the real cases shows slightly better results.  The results show that for most $ \beta $,  approximately $ N = 10 $ measurements is sufficient to achieve over $ 99 \% $ fidelity.

\begin{figure}[t]
    \centering
    \includegraphics[width=0.8\columnwidth]{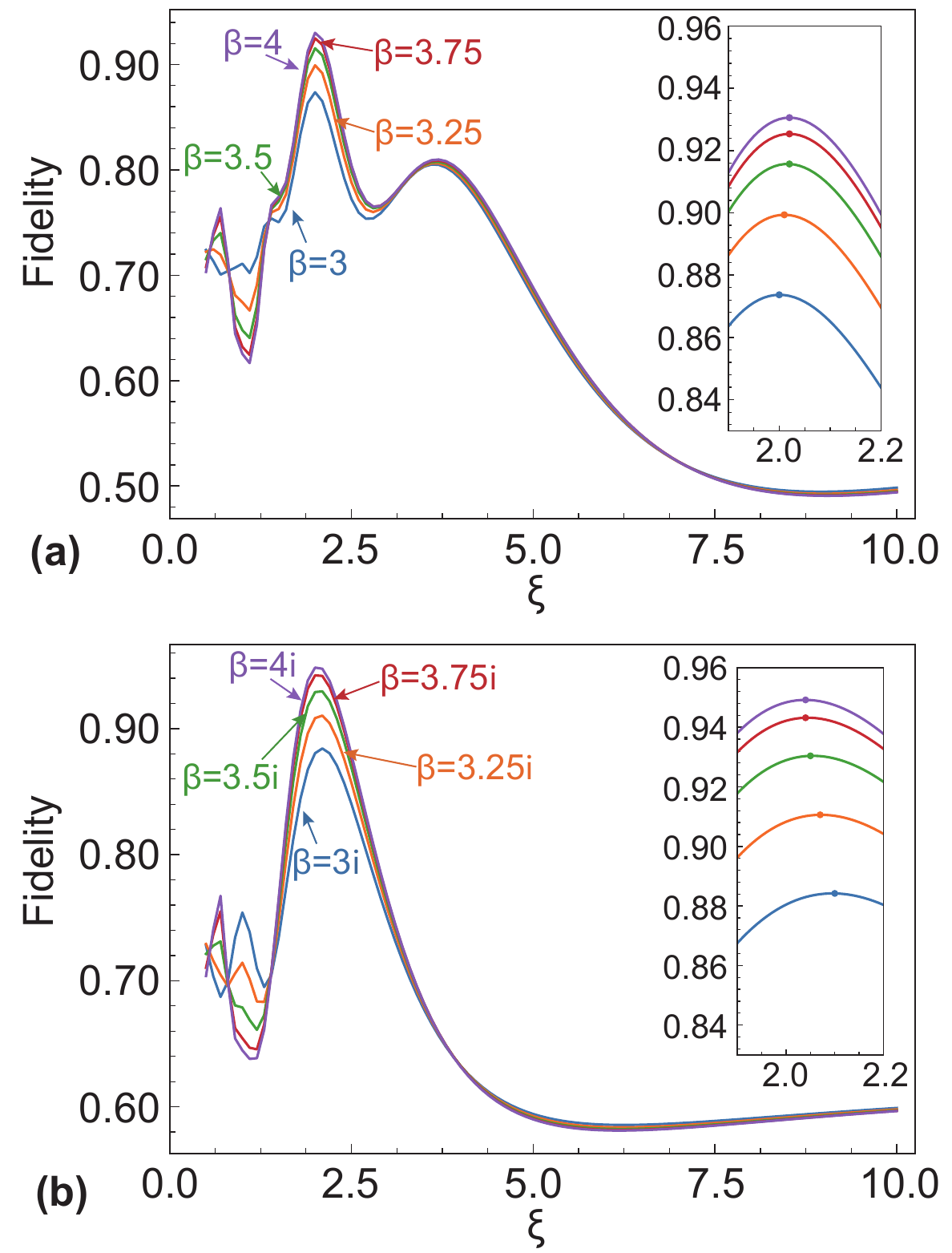}
    \caption{Average fidelity (\ref{avefidelity}) as a function of $\xi$ for Variant 1 teleportation with single-mode parity measurements. A zoomed-in plot is shown in the inset for better visibility of the optimal points. Parameters are \textbf{(a)} $ \beta \in \mathbb{R} $; \textbf{(b)} $ \beta \in \mathbb{I} $. Common parameters are $\mu=1/2, \nu=\sqrt{3}/2, \alpha=4, N=1, k_+=1, k_-=0, \tau = \pi/(\xi \bar{n} ) $.}
     \label{figure7}
\end{figure}

\subsection{Average over dispersive measurement outcomes}

Up to this point, we have only considered the measurement outcome $ \sigma_a' = + $, which experimentally corresponds to performing a postselection. In order to consider other outcomes, we now show that averaging over dispersive measurements still yields high fidelity results.  We still consider the $ \sigma_a = \sigma_b = + $ outcome, since all cases in (\ref{paritymeasstate}) are equivalent up to the correction.   

For the case that the $ a $ mode collapses to the state $ |0_L \rangle_{\bar{\chi}} $, the state on mode $ c $ is $ \mu | 0_L \rangle_\beta - \nu | 1_L \rangle_\beta$.  This requires a correction $ C_{++-} = Z_L $, which can be performed by applying (\ref{zlcorr}).   We calculate the average fidelity as 
\begin{align}
F^{\text{av}}_{\sigma_a \sigma_b } = \sum_{\sigma_a'= \pm} F_{\sigma_a \sigma_b \sigma_a'} p_{\sigma_a' |\sigma_a \sigma_b } , 
\label{avefidelity}
\end{align}
using the definitions (\ref{fiddef}) and (\ref{probdef}). 

Fig. \ref{figure7}(a) shows the average fidelity as a function of the time parameter $ \xi $. We observe a general behavior and performance that is similar to the postselected cases in Fig. \ref{figure5}.  However, the behavior at the optimal points differ slightly due to the averaging effects. For example, for $\beta= \alpha = 4$, we find that the best average fidelity is at the level of 93\%. A slightly higher fidelity level of $\sim 95\%$ can be obtained if the parameter $\beta$ is taken as $\beta=4i$, as seen in Fig. \ref{figure7}(b).

\section{Teleportation performance using joint parity measurements}
\label{jointparityresults}

In the previous section, we examined the performance of the teleportation using single-mode parity measurements. In this section, we examine the performance using joint parity measurements (Variant 2), as described in Sec. \ref{sec:joint}.  

As with the previous section, we start with the state (\ref{paritymeasstate}) and then perform the displacement $ D(\chi) $, followed by the joint parity measurement of modes $ a $ and $ b$.  We use the same parameters as we performed in Sec. \ref{singleparityresults} so that we can compare the results and determine which measurement scheme achieves superior fidelity. The postselected outcomes that we choose are for the case $\sigma_{ab} = +$. Here, we only present a few representative results for comparative analysis with the the single-mode parity measurement results.

Similarly to the definition in Sec. \ref{singleparityresults}, we now define the fidelity for the outcome $( \sigma_{ab}, \sigma_a') $ with joint parity measurements:
\begin{align}
F_{\sigma_{ab} \sigma_a'} =  &  ( \mu^* \langle 0_L |_\beta  +  \nu^* \langle 1_L | ) C_{\sigma_{ab} \sigma_a'}  \rho_{ \sigma_a' |\sigma_{ab} }  \nonumber \\
& \times C^\dagger_{\sigma_{ab} \sigma_a'} | ( \mu | 0_L \rangle_\beta  + \nu | 1_L \rangle_\beta )
\label{fidvariant2}
\end{align}
where 
\begin{align}
 \rho_{ \sigma_a' |\sigma_{ab} }  = \frac{\text{Tr}_{a,b} | \psi_{ \sigma_a' |\sigma_{ab}  } \rangle  \langle \psi_{ \sigma_a' | \sigma_{ab}} |}{p_{\sigma_a' | \sigma_{ab} } }
 \label{rhojoint}
\end{align}
is the normalized state on mode $ c $ tracing out modes $ a,b $. $ C_{\sigma_{ab} \sigma_a'}  $ is the logical correction operation that appears in Table \ref{tab:finalstatesjoint}. 
The final state after applying the displaced dispersive measurement is
\begin{align}
| \psi_{ \sigma_a' |\sigma_{ab}  } \rangle = \frac{M_{\sigma_a'} D(\chi) | \psi_{\sigma_{ab} } \rangle}{\sqrt{p_{\sigma_{ab}}  }}
\end{align}
where the conditional probability is
\begin{align}
p_{\sigma_a' | \sigma_{ab} }=  \langle \psi_{ \sigma_a' | \sigma_{ab} }   | \psi_{\sigma_a' | \sigma_{ab}  } \rangle . 
\end{align}

\begin{figure}
    \centering
    \includegraphics[width=0.8\columnwidth]{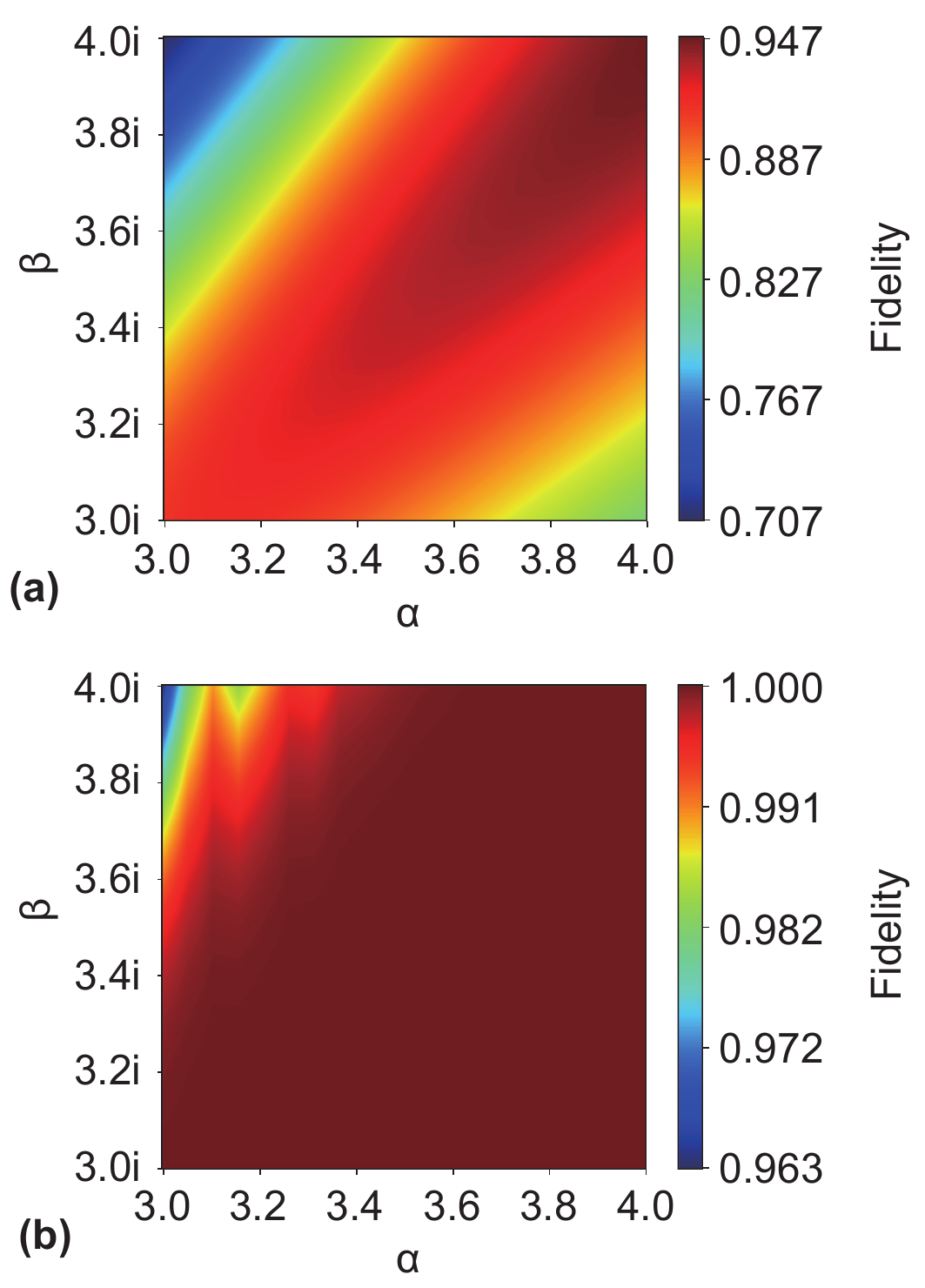}
    \caption{Fidelity (\ref{fidvariant2}) in the space of  $\alpha \in \mathbb{R} $ and $\beta \in \mathbb{I} $ for Variant 2 teleportation with joint parity measurements. Parameters are \textbf{(a)} $N=1, k_+=1, k_-=0 $; \textbf{(b)} $N=1000, k_+=1000, k_-=0 $. Common parameters are $ \mu=1/2, \nu=\sqrt{3}/2,  \tau = \pi/(4\alpha^2) $.  }
        \label{figure8}
\end{figure}

\subsection{$ \alpha \in \mathbb{R} $ and $ \beta \in \mathbb{I}  $ , $ N = 1$ measurements }
\label{singlesec:A}

In Fig. \ref{figure8}(a) we plot the fidelities for various choices of $ \alpha \in \mathbb{R} $ and $ \beta \in \mathbb{I}  $ with fixed interaction time $ \tau = \pi/(4 \alpha^2) $.  This is the equivalent of Fig. \ref{figure3}(a) but using joint parity measurements.  As before,  this interaction time is not the optimal time as derived in (\ref{optimaltime}) unless $ |\beta| = |\alpha |$. For the final dispersive measurement we consider the measurement outcome $ \sigma_a' = + $ corresponding to a collapse on the $ |\pm \chi \rangle $ states (see Fig. \ref{fig:coherent}(e)).   We again see high fidelities are obtained for various choices of $ \alpha, \beta $.  The best fidelities areas occur along the line $ \beta= i \alpha $, with the highest fidelities occurring for large $ |\alpha|, | \beta |$.  Generally, very similar results are obtained to the single-mode parity measurement case. We thus conclude that the effect on the two states (\ref{paritymeasstate}) and (\ref{jointparitymeasstate}) are generally similar. We give the further discussion of this in Appendix \ref{appendixB:part2}

\subsection{$ \alpha \in \mathbb{R} $ and $ \beta \in \mathbb{I}  $ , $ N = 1000$ measurements }

Fig. \ref{figure8}(b) shows the results for the same parameters as Fig. \ref{figure8}(a), but the dispersive measurements at the end are repeated $ N = 1000$ times. We again use the fixed interaction time $ \tau = \pi/(4 \alpha^2) $, which is suboptimal as derived in (\ref{optimaltime}) unless $ |\beta| = |\alpha |$.  We see that increasing the number of measurements produces an improvement of the fidelity in comparison to Fig. \ref{figure8}(a).  Again, very similar results are seen to the single-mode parity measurement case given in Fig. \ref{figure3}(b).

\begin{figure}
    \centering
    \includegraphics[width=0.8\columnwidth]{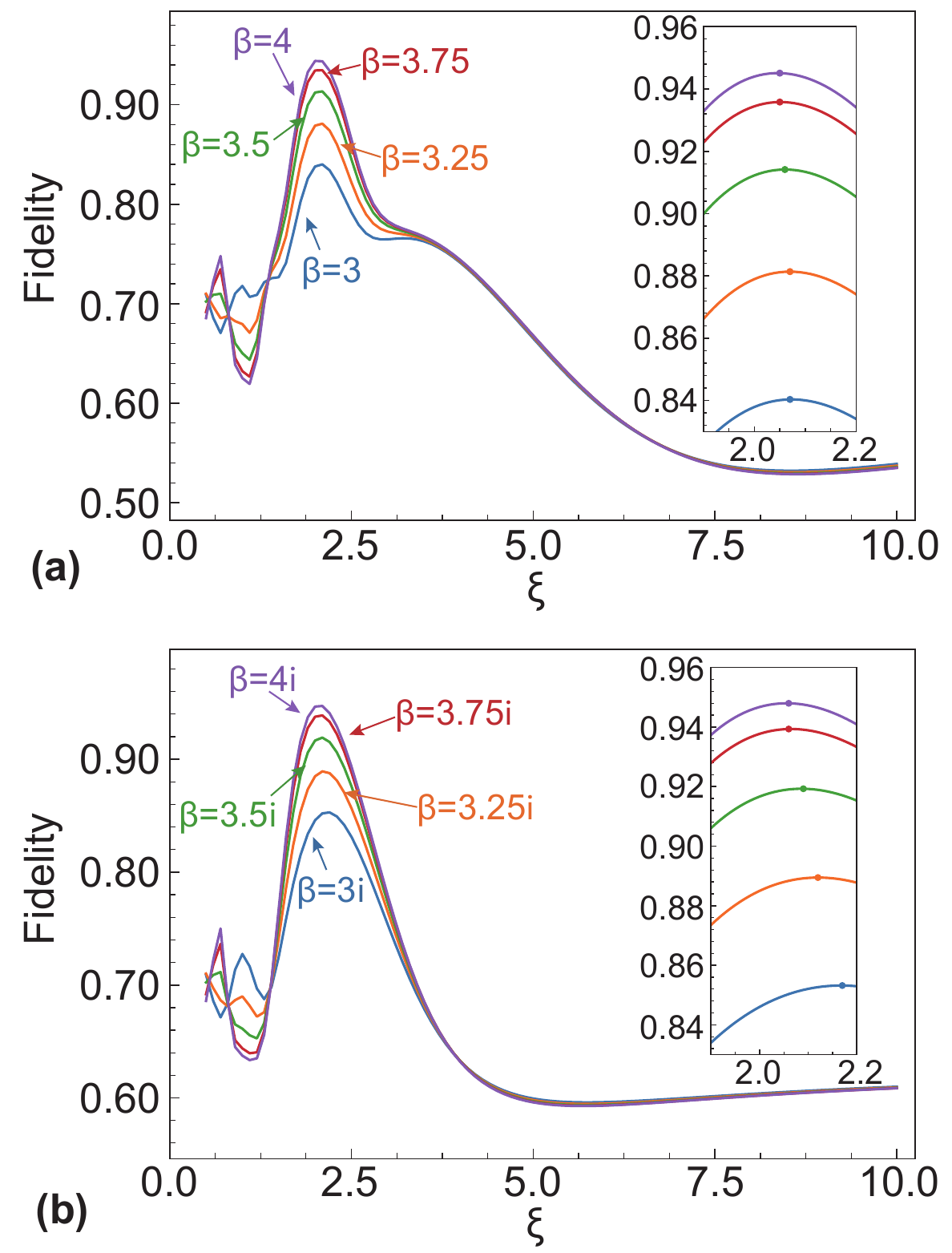}
    \caption{Fidelity (\ref{fidvariant2}) as a function of $\xi$ for Variant 2 teleportation with joint parity measurements. The insets show the same plot zoomed into the peak region.  Parameters are chosen such that \textbf{(a)} $ \beta \in \mathbb{R} $; \textbf{(b)} $ \beta \in \mathbb{I} $. Common parameters are $ \mu=1/2, \nu=\sqrt{3}/2, \alpha=4, N=1,  k_+=1, k_-=0, \tau = \pi/(\xi \bar{n} ) $. }        \label{figure9}
\end{figure}

\subsection{Dependence on the interaction time $\tau = \pi/(\xi \bar{n} )$}

Finally, we calculate the fidelity as a function of the interaction time under joint parity measurement. We observe similar behaviors for both $\beta \in \mathbb{R}$ and $\beta \in \mathbb{I}$ in Fig. \ref{figure9} compared to the single-mode parity measurement results of Fig. \ref{figure5}. In Fig. \ref{figure9}(a) we examine the $ \beta \in \mathbb{R}$ case.  For $ \beta < \alpha $, the fidelity achieved with the joint parity measurement method is slightly lower than that obtained with the single-mode parity measurement (see inset of Fig. \ref{figure9}(a)). However, when $\beta = \alpha$, both methods produce identical results. Further, Fig. \ref{figure9}(b) shows that the fidelities remain unchanged under joint parity measurement if $\beta \in \mathbb{I}$. We thus again conclude that in the joint measurement case, $ \tau = \pi/(2 \bar{n}) $ is close to the optimal time in all cases.  Thus (\ref{optimaltime}) is a universally good approximation to the optimal time for both $ \beta \in \mathbb{R},\mathbb{I}$, and for both single-mode and joint parity measurements.

\section{Decoherence}
\label{decoherence}

The study of decoherence is crucial for the realization of quantum teleportation, as any experiment is inevitably affected by the environment. In particular, for teleportation protocols based on 
Schr{\"o}dinger cat states, their macroscopic nature make them more sensitive to decoherence compared to other types of quantum states \cite{zurek2003decoherence}. In this section, we introduce decoherence into our protocol, and evaluate its performance to confirm that the protocol remains practical within experimentally achievable limits. 

Here, we consider the loss channel, which can be characterized by the a set of Kraus operators \cite{nielsen2010quantum}
\begin{align}
A_k = \sum_{n=k}^\infty \sqrt{n \choose k} \sqrt{\gamma}^{n-k} \sqrt{1-\gamma}^{k} |n-k\rangle \langle k|,
\end{align}
where $1-\gamma$ is the loss probability, $k$ is the number of lost bosons. Suppose that the loss occurs on mode $a$, affecting the ability to distinguish between $\chi$ and $\bar{\chi}$ effectively. Note that we only discuss the case of Variant 1, as the results obtained for Variant 2 are similar. We apply $A_k$ to (\ref{paritymeasstate}) on mode $a$ such that the state is
\begin{align}
| \psi^{k}_{\sigma_a \sigma_b } \rangle = \frac{A_k | \psi_{\sigma_a \sigma_b } \rangle}{\sqrt{p_k} \sqrt{p_{\sigma_a \sigma_b} }},
\end{align}
where
\begin{align}
p_k = \langle \psi_{\sigma_a \sigma_b }|A^{\dagger}_k A_k |\psi_{\sigma_a \sigma_b }\rangle.
\end{align}
is the probability that $k$ bosons are lost. 
We now define the fidelity for the measurement outcome $( \sigma_a, \sigma_b,  \sigma_a') $ as 
\begin{align}
F^{k}_{\sigma_a \sigma_b \sigma_a'} =  &  ( \mu^* \langle 0_L |_\beta  +  \nu^* \langle 1_L | ) C_{\sigma_a \sigma_b \sigma_a'}   \rho^{k}_{ \sigma_a'| \sigma_a \sigma_b}  \nonumber \\
& \times C^\dagger_{\sigma_a \sigma_b \sigma_a'} | ( \mu | 0_L \rangle_\beta  + \nu | 1_L \rangle_\beta ) ,
\end{align}
where
\begin{align}
 \rho^{k}_{ \sigma_a'| \sigma_a \sigma_b}  = \frac{\text{Tr}_{a,b} | \psi^{k}_{ \sigma_a' | \sigma_a \sigma_b} \rangle  \langle \psi^{k}_{\sigma_a' | \sigma_a \sigma_b  } |}{p^{k}_{\sigma_a' | \sigma_a \sigma_b } }
\end{align}
is the state on mode $c$ after tracing out modes
$a$ and $b$. $ C_{\sigma_a \sigma_b \sigma_a'}  $ is the same correction operation as we introduced in Sec. \ref{singleparityresults} and \ref{jointparityresults}. The state after applying the displacement and dispersive measurement is
\begin{align}
| \psi^{k}_{\sigma_a' | \sigma_a \sigma_b  } \rangle = M_{\sigma_a'} D(\chi) | \psi^{k}_{\sigma_a \sigma_b } \rangle,
\end{align}
and the conditional probability is
\begin{align}
p^{k}_{\sigma_a' | \sigma_a \sigma_b }=  \langle \psi^{k}_{\sigma_a' | \sigma_a \sigma_b  }   | \psi^{k}_{\sigma_a' | \sigma_a \sigma_b  } \rangle . 
\label{probdef}
\end{align}
We then obtain the fidelity averaged over $k$
\begin{align}
F^{\text{av}}_{\sigma_a \sigma_b \sigma_a'} = \sum_k p_k F_{\sigma_a \sigma_b \sigma_a'}^k.
\label{avefidelityloss}
\end{align}
\begin{figure}
    \centering
    \includegraphics[width=\columnwidth]{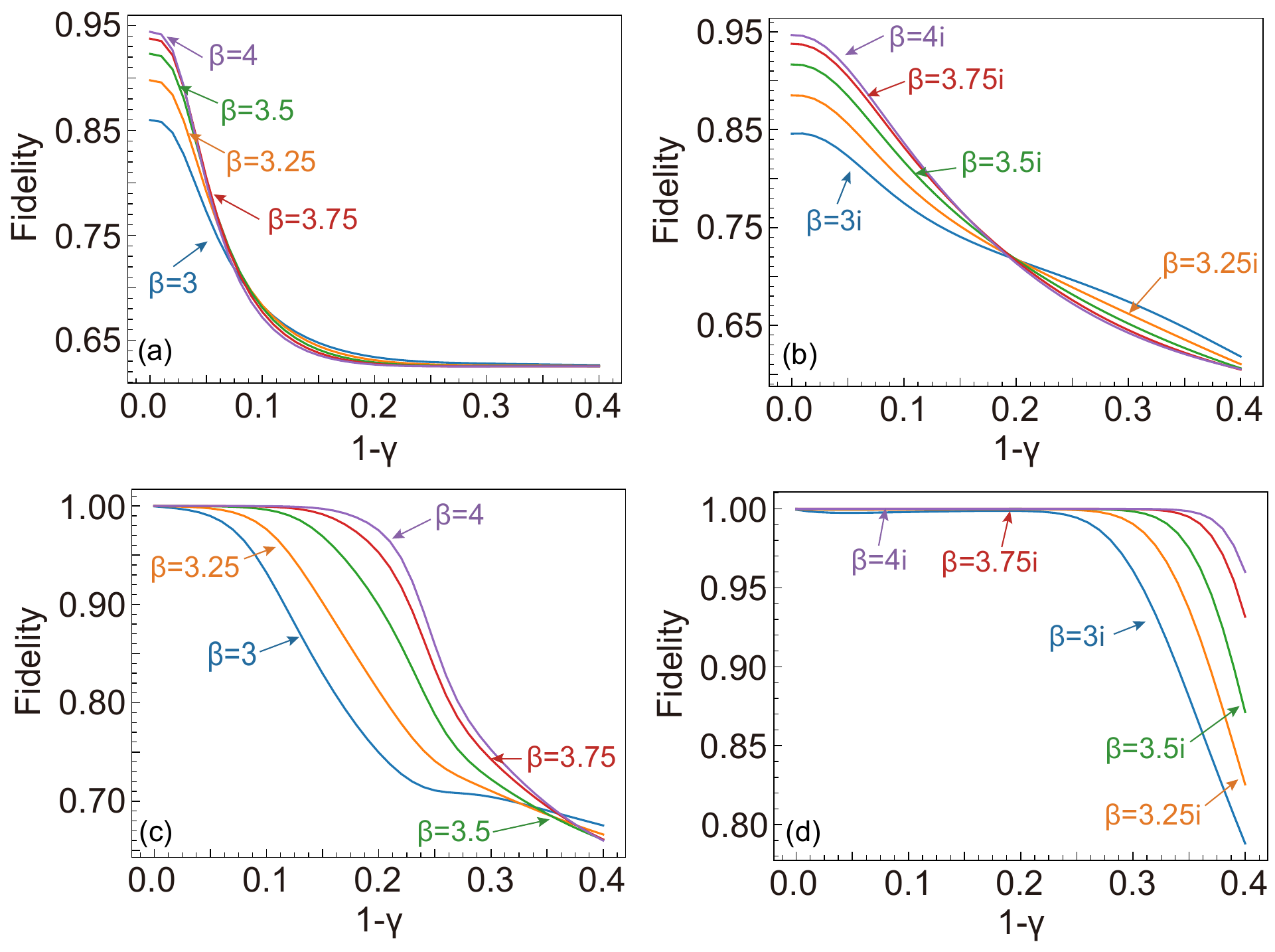}
    \caption{Fidelity (\ref{avefidelityloss}) as a function of $1-\gamma$ for Variant 1 teleportation with single-mode parity measurements.  Parameters are chosen such that \textbf{(a)} $ \beta \in \mathbb{R}, N=1,  k_+=1, k_-=0 $; \textbf{(b)} $ \beta \in \mathbb{I}, N=1,  k_+=1, k_-=0 $; \textbf{(c)} $ \beta \in \mathbb{R}, N=1000,  k_+=1000, k_-=0$; \textbf{(d)} $ \beta \in \mathbb{I}, N=1000,  k_+=1000, k_-=0 $. Common parameters are $ \mu=1/2, \nu=\sqrt{3}/2, \alpha=4, \tau = \pi/(2 \bar{n} ) $. }        \label{figuredecoherence}
\end{figure}

Fig. \ref{figuredecoherence} shows the average fidelity (\ref{avefidelityloss}) versus the probability of loss $1-\gamma$ under single-mode measurement for various parameters. First, we see that for loss probabilities in the range of a few $ \sim \% $, the fidelities are largely unchanged.  For larger loss probabilities, we observe that the larger $|\beta|$ of cat states have a faster decoherence rate, which is expected as larger cat states are more susceptible to loss.  We see that the fidelity converges with loss rate for $\beta \in \mathbb{R}$ (Fig. \ref{figuredecoherence}(a)). This is mainly because the ability of distinguishing the distribution no longer changes significantly. However, for $\beta \in \mathbb{I}$ the fidelity seems to converge at a higher loss rate (Fig. \ref{figuredecoherence}(b)). For example, the fidelity within the choice of parameters $\beta \in \mathbb{I}$ is significantly higher than that using $\beta \in \mathbb{R}$ for the plotted range. Hence, we conclude that the protocol is more robust in this case.

In previous sections, we showed that by using repeated dispersive measurements, one can obtain a fidelity approaching 1. Similarly, we now examine the multiple measurements case in the presence of loss. The results are shown in Fig. \ref{figuredecoherence}(c)(d). We find that compared to single measurements, the fidelity is substantially improved, especially for larger choices of the amplitude $|\beta|$. For example, when $|\beta| = 4$, the system can almost completely resist loss up to probabilities $1- \gamma < \sim 15\%$, as illustrated in Fig. \ref{figuredecoherence}(c). Likewise, Fig. \ref{figuredecoherence}(d) shows that choosing $\beta \in \mathbb{I}$ leads to significantly more robust results. Choosing $\beta = 4i$ and performing multiple measurements can yield a nearly perfect fidelity, even the probability of loss is up to 35\%.

\section{Conclusions}
\label{conclusion}

We have introduced a protocol to perform the teleportation of cat state encoded qubits using beam splitter operations and dispersive measurements.  Our protocol is an alternative to the standard CV teleportation protocol, and is preferable when quadrature measurements are not easily performed experimentally. Such systems include those that have an engineerable Jaynes-Cummings Hamiltonian, such as circuit-QED and -QAD systems, and trapped ions, where the types of operations that we have used in the teleportation protocol have been demonstrated in several contexts.   At first glance our protocol, summarized in Fig. \ref{fig:circuit}, is reminiscent to standard CV teleportation.  However, it differs in the type of measurements that are performed, and the different nature of the states that occur throughout. Parity measurements only partially collapse the wavefunction of a bosonic mode and this results in an additional measurement being required to complete the teleportation.  We showed that by using repetitive dispersive measurements it is possible to perform this collapse with high fidelity.  Even at the single measurement level, fidelities in the vicinity of $ \sim 95 \% $ can be attained, which far exceeds the classical bounds for qubit teleportation, at $ 66 \% $ \cite{massar1995optimal}.  

We introduced two variants of the teleportation protocol, using either two single-mode parity measurements (Variant 1) or a joint parity measurement (Variant 2).  Both variants work with similar performance and have a similar dependence with parameters.  We also considered parameter choices with real or imaginary choices of $ \beta $ and showed that the performance was similar.  Joint parity measurements are attractive from the standpoint of the optimality of the number of measurement outcomes.  We demonstrated that the displaced dispersive measurement procedure, where the total number of measurements plays a crucial role and can lead to a near-perfect fidelity. Our simulation results show that by choosing the proper optimal interaction time (\ref{optimaltime}) and the absolute eigenvalues $|\alpha|,|\beta|$, one can obtain a high level of fidelity.  In terms of the protocol itself, a larger $|\alpha|,|\beta|$ gives better results since it satisfies the approximation (\ref{approxrelation}).  When including decoherence, selecting larger amplitude of cat states and performing multiple measurements can obtain near-perfect fidelity, despite larger amplitude cat states being more sensitive to loss \cite{zurek2003decoherence,byrnes2021quantum}. Given that manipulations of cat states are already an experimental reality \cite{Wang2016jointparity,bild2023schrodinger}, and the schemes of stabilization of cat states have been proposed \cite{Zapletal22PRXQ, Marquet24prx}, we believe that our protocol is within the realm of practical implementation.

\begin{acknowledgments}
TB is supported by the SMEC Scientific Research Innovation Project (2023ZKZD55); the Science and Technology Commission of Shanghai Municipality (22ZR1444600); the NYU Shanghai Boost Fund; the China Foreign Experts Program (G2021013002L); the NYU-ECNU Institute of Physics at NYU Shanghai; the NYU Shanghai Major-Grants Seed Fund; and Tamkeen under the NYU Abu Dhabi Research Institute grant CG008.  MF was supported by the Swiss National Science Foundation Ambizione Grant No. 208886, and by The Branco Weiss Fellowship -- Society in Science, administered by the ETH Z\"{u}rich.
\end{acknowledgments}

\appendix

\section{Additional simulation results for single-mode parity measurements  }
\label{appendixA}

In this section, we provide additional simulation results for Variant 1 teleportation with single-mode parity measurements. We offer a more detailed explanation of the occurrence of different optimal fidelities.

\subsection{$ \alpha, \beta \in \mathbb{R} $,  $ \tau = \pi/(2 \bar{n} ) $, $ N = 1$ measurements}
\label{appendixA:part1}

In Sec. \ref{sec:taudependence}, we verified that $ \tau = \pi/(2 \bar{n} ) $ gives a good estimate of the optimal time.  By analyzing how the state distribution varies with different parameters, we can gain a more intuitive understanding of the factors that affect the fidelity.  In order to check that reasonable projections are being performed to distinguish between the $ |\pm \chi \rangle $ and $ |\pm \bar{\chi} \rangle $ states (see Fig. \ref{fig:coherent}), it is illuminating to plot the Fock basis distributions.

\begin{figure}[t]
    \centering
    \includegraphics[width=0.8\columnwidth]{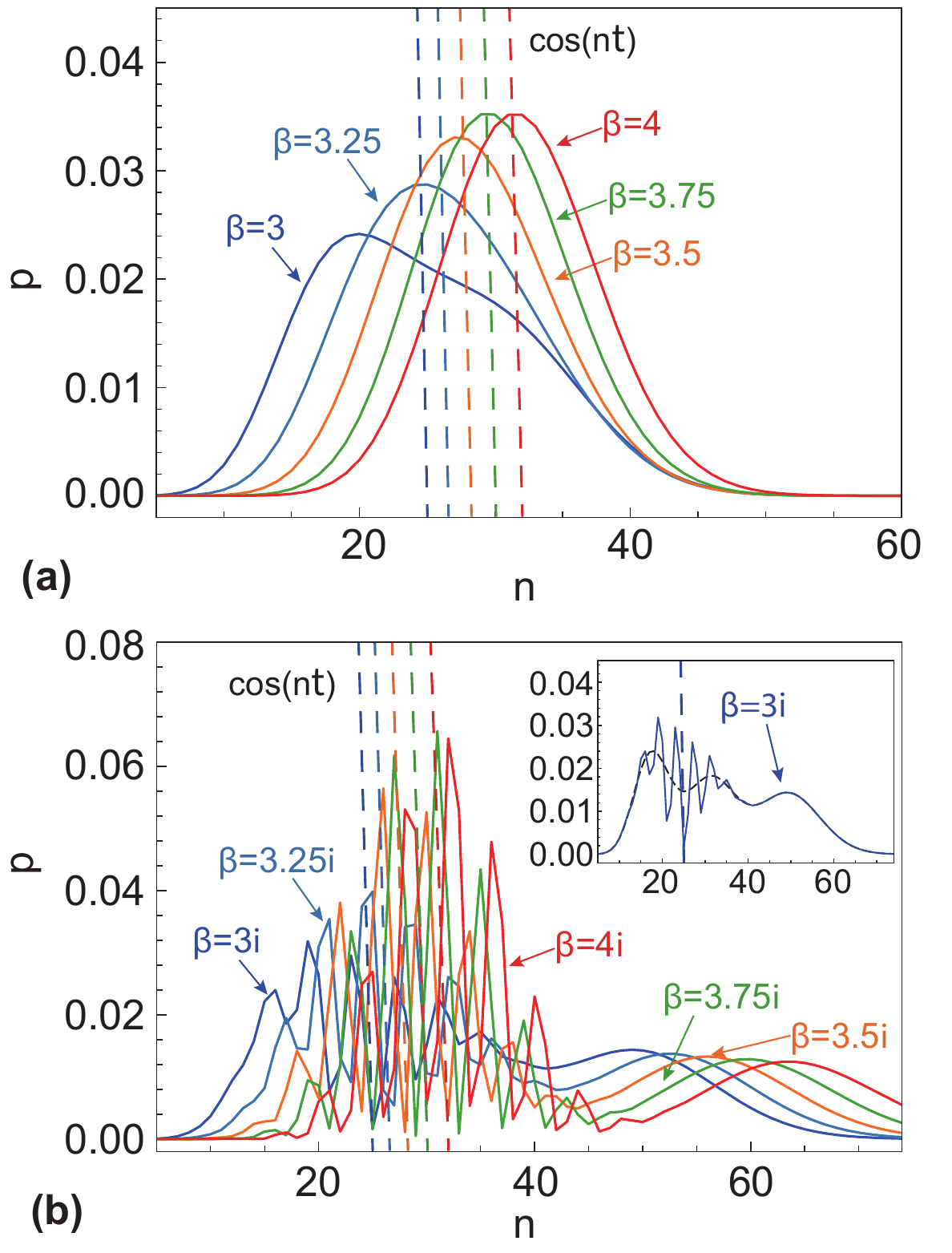}
    \caption{The probability distributions in the Fock basis $ p_n = \langle n | \rho_{\sigma_a \sigma_b} | n \rangle  $ for the single-mode parity measured states $ \rho_{\sigma_a \sigma_b} = \text{Tr}_{b,c} D(\chi) | \psi_{\sigma_a \sigma_b} \rangle \langle  \psi_{\sigma_a \sigma_b} | D^\dagger (\chi) /p_{\sigma_a \sigma_b} $.  Parameters are chosen such that \textbf{(a)} $ \beta \in \mathbb{R} $; \textbf{(b)} $  \beta \in \mathbb{I}  $.  The inset shows detail of the  $\beta=3i$ case.  The dotted lines in the inset show the averaged curve.  The dashed lines are $\cos (n \tau) $ for different values of $\beta$ respectively, using $ \tau = \pi/(2\bar{n}) $.  The lines appear as nearly vertical because the scale of the plot is much smaller than the full range of the cosine function $ [-1,1] $.  Common parameters are $\mu=1/2, \nu=\sqrt{3}/2,  \alpha = 4, N=1,  k_+=1, k_-=0 , \tau = \pi/(2 \bar{n})$.}
    \label{figurenew-1}
\end{figure}

\begin{figure}[t]
    \centering
    \includegraphics[width=0.8\columnwidth]{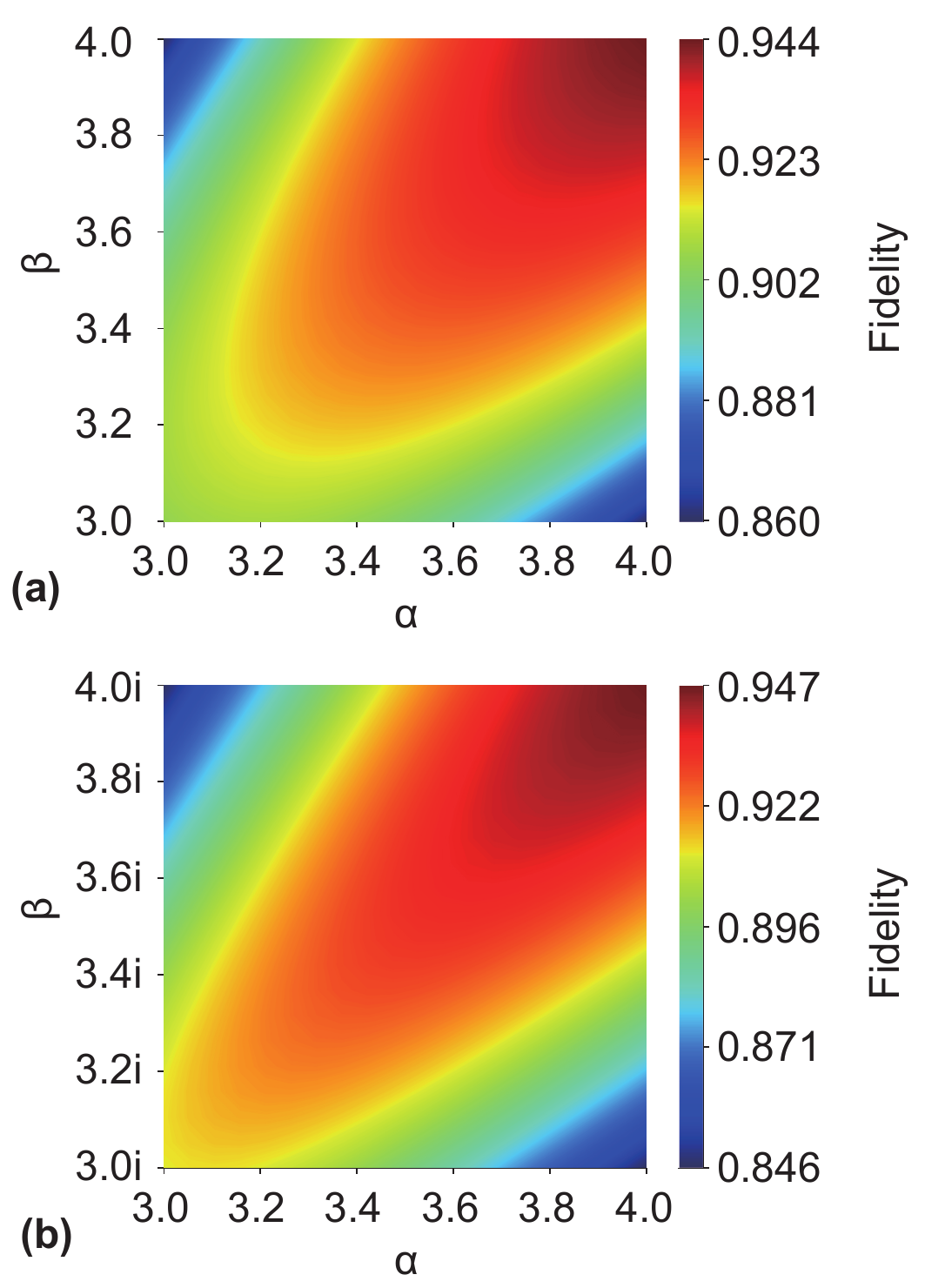}
    \caption{Fidelity (\ref{fiddef}) in the space of  $\alpha$ and $\beta$ for Variant 1 teleportation with single-mode parity measurements. We choose parameters such that \textbf{(a)} $\alpha $,$ \beta \in \mathbb{R} $; \textbf{(b)} $\alpha \in \mathbb{R} $ and $ \beta \in \mathbb{I}  $.  Common parameters are $\mu=1/2, \nu=\sqrt{3}/2, N=1, k_+=1, k_-=0, \tau = \pi/(2 \bar{n})$.  }
    \label{densityfinal-3}
\end{figure}

The Fock basis distributions of the states $D(\chi) | \psi_{\sigma_a \sigma_b} \rangle $ on mode $a$ for various choices of $ \alpha, \beta \in \mathbb{R} $ are plotted in Fig. \ref{figurenew-1}(a).  We show the probabilities in the region of the $|\pm \bar{\chi} \rangle$ peak (see Fig. \ref{fig:coherent}(f)).  The peaks $|\pm \chi \rangle$ are out of range of the plots and are not visible. We note that the states on mode $a$ are now in a superposition of the $|\pm \bar{\chi} \rangle$ states, hence the actual probabilities are not the ideal Poissonian distributions as plotted in Fig. \ref{fig:coherent}(f).  For the various values of $ \beta $, we see that the zero of the cosine function coincides with the peak of the probabilities.  We thus see that  $ \tau = \pi/(2 \bar{n} ) $ is indeed a nearly optimal choice for removing the distribution.   The peak of the distributions shifts to the left with decreasing $|\beta|$ and also broadens. The broadening reduces the effectiveness of the cosine function to remove this peak, leading a relatively lower fidelity. This can be seen in Fig. \ref{densityfinal-3}(a), where the superior fidelities are obtained near the top right corner of the plot. We also observe that this plot exhibits a symemtry along the diagonal line $ \beta = \alpha $ as both the interaction time (\ref{optimaltime}) and the distribution $p_n$ themselves possess a symmetry between $\alpha$ and $\beta$. For the single measurement case $N=1$ with outcome $ \sigma_a' = + $,  we find a good fidelity of $\sim 94\%$ can be obtained (see Fig. \ref{densityfinal-3}(a)). We expect higher fidelities with larger $|\alpha|$ and $|\beta|$.

\subsection{$ \alpha \in \mathbb{R} $ and $ \beta \in \mathbb{I}  $ , $ \tau = \pi/(2 \bar{n}) $, $ N = 1 $ measurements}
\label{appendixA:part2}

We now consider the equivalent case as Sec. \ref{appendixA:part1}, but with $ \beta \in \mathbb{I}  $. The Fock basis distributions on mode $ a $ of the state  $D(\chi) | \psi_{\sigma_a \sigma_b} \rangle $ are shown in Fig. \ref{figurenew-1}(b). Due to the superposition of the $| \pm \bar{\chi} \rangle$ states, the distribution has some interference effects, giving rise to oscillatory behavior (see inset of Fig. \ref{figurenew-1}(b) for further detail).   Further, unlike Fig. \ref{figurenew-1}(a), here the $|+\chi\rangle$ state lies closer to the $| \pm \bar{\chi} \rangle$ states, resulting in a significant overlap.  This can be also seen in Fig. \ref{fig:coherent}(e), where the $|+\chi\rangle$ peak is closer to the  $| \pm \bar{\chi} \rangle$ peak.  This potentially impacts the fidelity especially for small values of $ |\alpha|, |\beta |$. This behavior accounts for the sensitive optimal points seen in Fig. \ref{figure5}(b). The interference effects and the presence of the $|+\chi\rangle$ peak can shift the distribution peak, which changes the optimal interaction time. When $\beta = 4i$, the ability to distinguish between two states is maximized, and with $\xi \approx 2.06$ with a fidelity level of 95\% is reached, as shown in Fig. \ref{figure5}(b). The fidelities for various $ \alpha, \beta $ are shown in Fig. \ref{densityfinal-3}(b). We first notice that this is also a symmetric pattern, with the better fidelity areas in the top right corner. Moreover, the fidelity are generally slightly lower than the $ \beta \in \mathbb{R} $.  Despite this, the highest fidelity in fact exceeds the $ \beta \in \mathbb{R} $ case, and excellent fidelities are obtained particularly for $ |\beta | = |\alpha |$.   Again, we expect further improvements of the fidelity with larger $ |\alpha |, |\beta |$.

\begin{figure}[t]
    \centering
    \includegraphics[width=0.8\columnwidth]{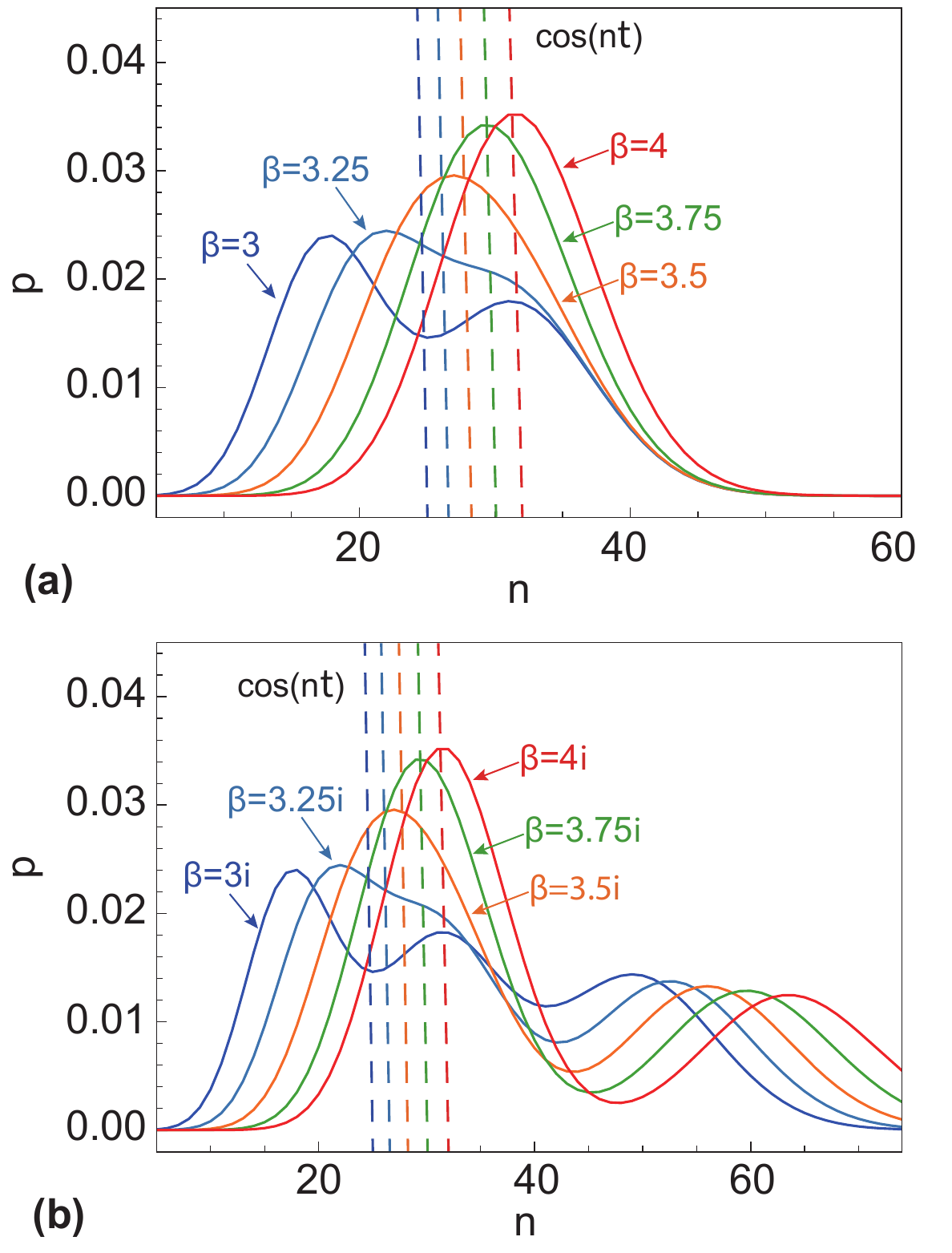}
    \caption{The probability distributions in the Fock basis $ p_n = \langle n | \rho_{\sigma_{ab}} | n \rangle  $ for the joint parity measured states $ \rho_{\sigma_{ab}} = \text{Tr}_{b,c} D(\chi) | \psi_{\sigma_{ab}} \rangle \langle  \psi_{\sigma_{ab}} | D^\dagger (\chi) /p_{\sigma_{ab}} $.   Parameters are chosen such that \textbf{(a)} $ \beta \in \mathbb{R} $; \textbf{(b)} $  \beta \in \mathbb{I}  $.   The dashed lines are $\cos (n \tau) $ for different values of $\beta$ respectively, using $ \tau = \pi/(2\bar{n}) $.  The lines appear as nearly vertical because the scale of the plot is much smaller than the full range of the cosine function $ [-1,1] $. Common parameters are $\mu=1/2, \nu=\sqrt{3}/2, \alpha = 4, N=1, k_+=1, k_-=0 , \tau = \pi/(2 \bar{n})$.}
    \label{figurenew-2}
\end{figure}

\section{Additional simulation results for joint parity measurements}

Here we perform the joint parity measurement version of Sec. \ref{appendixA} for Variant 2 teleportation.

\subsection{$ \alpha, \beta \in \mathbb{R} $,  $ \tau = \pi/(2 \bar{n} ) $, $ N = 1$ measurements}

\begin{figure}
    \centering
    \includegraphics[width=0.8\columnwidth]{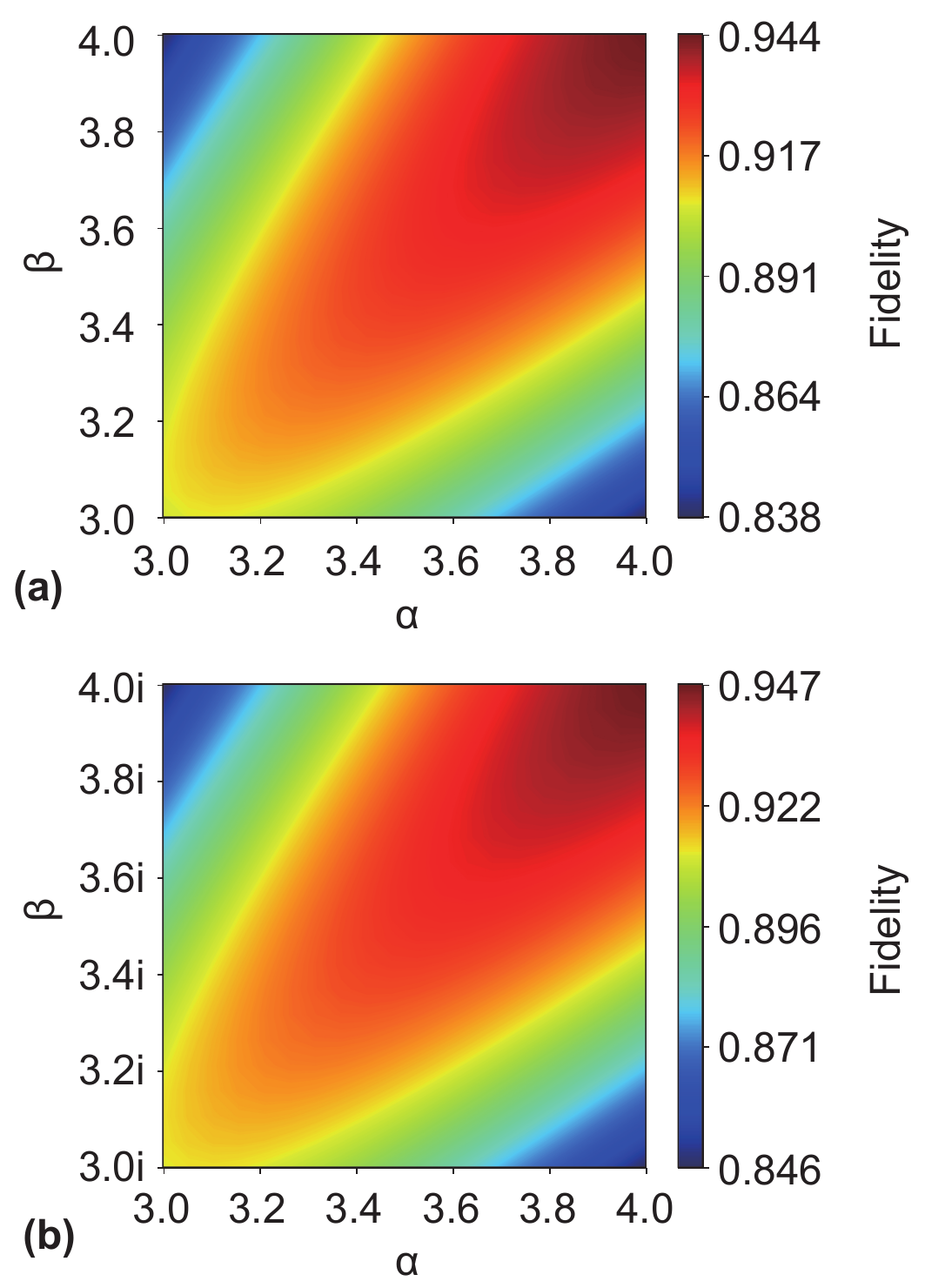}
    \caption{Fidelity (\ref{fidvariant2}) in the space of  $\alpha \in \mathbb{R} $ and $\beta \in \mathbb{I} $ for Variant 2 teleportation with joint parity measurements.  Parameters are \textbf{(a)} $\alpha $,$ \beta \in \mathbb{R} $; \textbf{(b)} $\alpha \in \mathbb{R} $ and $ \beta \in \mathbb{I}  $.  Common parameters are $\mu=1/2, \nu=\sqrt{3}/2, N=1, k_+=1, k_-=0, \tau = \pi/(2 \bar{n})$.  }
    \label{densityfinal-4}
\end{figure}

This section provides the analysis for the joint measurement version of Sec. \ref{appendixA:part1}.  We utilize the interaction time (\ref{optimaltime}) and show our results in Fig. \ref{densityfinal-4}(a). We find a some small differences between joint parity measurement and single-mode parity measurement in Fig. \ref{densityfinal-3}(a). The fidelity is decreased by $ \sim 2 \%$  relative to the single-mode parity measurement when $\beta=3$. Subsequently, the difference diminishes and eventually vanishes for larger $ \alpha, \beta $.  As can be seen from the distributions of $D(\chi) | \psi_{\sigma_{ab}} \rangle $ on mode $a$ in Fig. \ref{figurenew-2}(a), when $ \beta $ and $ \alpha $ are not equal, the distributions has a larger deviation from a Gaussian in comparison to Fig. \ref{densityfinal-3}(a).  As such, the cosine function is less effective at diminishing the distribution, giving rise to a relatively lower fidelity.  However, with larger $\beta$, the distributions of both two types of measurements become more similar. This can account for the fidelities of the two cases attaining the same value of $\sim 94.4\%$ when $\beta=\alpha = 4$.

\subsection{$ \alpha \in \mathbb{R} $ and $ \beta \in \mathbb{I}  $ , $ \tau = \pi/(2 \bar{n}) $, $ N = 1 $ measurements}
\label{appendixB:part2}

This section provides the analysis for the joint measurement version of Sec. \ref{appendixA:part2}. We utilize the interaction time (\ref{optimaltime}) and show our results in Fig. \ref{densityfinal-4}(b). We see virtually identical results to the single-mode measurement counterpart, Fig. \ref{densityfinal-3}(b).  Examining the distributions of the Fock states in Fig. \ref{figurenew-2}(b), interestingly, we see a rather different distribution to the single-mode counterpart, Fig. \ref{figurenew-1}(b).  Here, the interference patterns are all removed, and we have a smooth variation with $ n $.  Closer inspection shows that if one removes the interference by averaging them out, similar curves are obtained.  Since the distributions are averaged over via the trace in (\ref{rhojoint}), such interference patterns play a minimal role in the fidelity,  giving rise to the similar fidelities in Fig. \ref{densityfinal-4}(b).

	% How to do the references:
	%% 1) First uncomment the below and compile
\bibliographystyle{apsrev}
\bibliography{ref}
	%% 2) Copy the .bbl file to below and comment out the above two lines.
	%\begin{thebibliography}{28}

\end{document}